\documentclass[11pt,letterpaper]{article}
\usepackage[margin=0.5in]{geometry}
\usepackage{amssymb}
\usepackage{amsmath}
\usepackage{pifont}
\usepackage{array}
\usepackage{wrapfig}
\usepackage{graphicx}
\usepackage{url}
\usepackage{cite}
\usepackage{caption}
\usepackage{subcaption}
\usepackage{tablefootnote}

\usepackage[labelfont=bf,figurename=Fig]{caption}

\title{Fast and low-energy approximate full adder based on FELIX logic\thanks{Preprint Submitted to arXiv}}

\author{Seyed Erfan Fatemieh$^{1,}$\thanks{Corresponding author: Seyed Erfan Fatemieh (erfanfatemieh@eng.ui.ac.ir)}~, Samane Asgari$^{1,}$\thanks{Samane Asgari (s.asgari@eng.ui.ac.ir)}~ and Mohammad Reza Reshadinezhad$^{1,}$\thanks{Corresponding author: Mohammad Reza Reshadinezhad (m.reshadinezhad@eng.ui.ac.ir)}}

\begin{document}
\maketitle

{\noindent$^{1}$ \small \textit{Department of Computer Architecture, Faculty of Computer Engineering, University of Isfahan, Isfahan 8174673441, Iran}}
\section*{Abstract}
Nowadays, much data must be processed and moved between processing and memory units. New technologies and architectures have emerged to improve system performance and overcome the memory bottleneck. The memristor is a technology with both computing and memory capabilities. In-Memory Computing (IMC) can be performed by applying memristors to stateful design methods. The Fast and Energy-Efficient Logic in Memory (FELIX) logic is a stateful implementation logic. The way computations are performed can be changed to improve performance. Approximate design methods can be applied in error-resilient applications. In this paper, an approximate full adder circuit with exact $C_{out}$ and approximate $Sum$ outputs has been proposed using the FELIX design method for IMC in two different implementation approaches. The memristor count in the proposed FELIX-based Approximate Full Adder (FAFA) in the two proposed implementation approaches is improved by 14.28\% and 28.57\%, energy consumption is improved by 73.735\% and 81.754\%, and the number of computational steps in both approaches is improved by 66.66\% compared to the exact FELIX-based full adder. Two different scenarios are considered for evaluating the FAFA. The results of error analysis and evaluations of the FAFA in three different image processing and average pooling applications confirmed that FAFA has high accuracy and acceptable performance. 

\subsection*{Keywords}
Memristor, Approximate Computing, Approximate Full Adder, In-Memory Computing, Image Processing.

\section{Introduction} \label{sec1}
Recently, numerous efforts have been undertaken to address the Von-Neumann bottleneck or memory wall. This bottleneck arises from the significant energy consumption and delays in transferring data between memory and processing units \cite{ref1}. Most of the computational energy comes from the energy required to move data between these units because the energy to access memory increases dramatically along the memory hierarchy (from cache to off-chip memory) \cite{ref2,ref3}. Pedram et al. \cite{ref4} have stated that the energy consumption of a 16-bit adder in 45 nm Complementary Metal-Oxide Semiconductor (CMOS) technology is 3556 times less than the energy needed to access Dynamic Random Access Memory (DRAM). Also, the DRAM access time is 25 times higher than the 32-bit adder delay in CMOS technology \cite{ref2}. This problem significantly reduces system performance in applications where the volume of data is large and calculations are complex (e.g., machine learning, neural networks, video, and image processing). The physical proximity of memory and processing units was proposed to reduce energy consumption and data movement delay \cite{ref5}. However, this solution is only able to eliminate data movement partially. On the other hand, reducing the size of MOS Field Effect Transistors (MOSFETs) has faced limitations. These limitations include reduced reliability and gate control, increased subthreshold leakage currents, leakage power, and a cost wall \cite{ref6,ref7,ref8}. Consequently, alternative technologies are required to overcome the above problems. Research has been carried out in the past decades, which resulted in a shift towards emerging technologies such as Carbon Nanotube FET (CNFET), Quantum dot Cellular Automata (QCA), Ferroelectric FET (FEFET), nanomagnet logic, and memristors \cite{ref3,nref2}.

Memristors can have computing and data storage capabilities so that memristive crossbar arrays can enable In-Memory Computing (IMC) \cite{ref10,ref11}. IMC carries out fundamental computational tasks directly within the memory array, eliminating the need to transfer data to a separate processing unit \cite{ref3,ref12}. This type of operation brings benefits such as saving energy for reading, writing, and transferring data \cite{ref13}. Stateful design methods such as Material Implication (IMPLY), Fast and Energy-Efficient Logic in Memory (FELIX), and Memristor-Aided loGIC (MAGIC) are compatible with the crossbar array structures. Memristors offer many advantages, such as high scalability, compatibility with CMOS devices, low power consumption \cite{ref14,ref15}, high switching speed and low switching energy \cite{ref16}, and non-volatile behavior. Its non-volatile behavior means that if the power supply is turned off and turned on after a long period, the memristor maintains its last resistance value until a current or voltage is applied to it again \cite{ref17}. Memristors can be applied in various fields such as generators and filters \cite{ref18,ref19}, neural networks \cite{ref20}, and logic circuits \cite{ref1,ref21}.

To enhance performance, the methods for performing computations can be adapted in conjunction with technological advancements. In error-resilient applications, approximate computing can be used. Approximate computing brings advantages such as reducing area, energy consumption, delay, and hardware complexity \cite{ref1,ref22}. Achieving these benefits comes at the cost of losing an acceptable amount of accuracy. Therefore, approximate computing can only be applied in applications that do not need 100\% precision, including machine learning, image and video processing, and pattern recognition \cite{ref22}. Designing approximate circuits is possible in several ways. Certain components of exact circuits can be removed or simplified, and some values in their truth table can be modified to reduce hardware complexity \cite{ref22}. When the approximate computing method is applied, criteria such as Error Rate (ER), Error Distance (ED), Mean Error Distance (MED), and Normalized MED (NMED) should be used along with circuit evaluation criteria such as energy consumption and delay, to obtain a comprehensive evaluation of the proposed circuit performance.

Adders and multipliers are applied in almost every arithmetic calculation in a computer system \cite{ref13}. Therefore, because the performance of the systems is influenced by the wide applicability of these units in numerous applications and the high volume of data required for processing, an approximate full adder circuit for IMC has been proposed in this paper. The proposed FELIX-based Approximate Full Adder (FAFA) can be used well in error-resilient applications since the erroneous bit is not propagated from the Least Significant Bits (LSBs) structure to the Most Significant Bits (MSBs). The FAFA has been designed using the FELIX logic method, which is compatible with the structure of crossbar arrays. The key contribution of this paper is proposing a high-speed approximate full adder implementation algorithm (FAFA) with low energy consumption applicable in IMC alongside presenting an approximate truth table with exact $C_{out}$. Optimizing the execution algorithm of the $1^{st}$ implementation approach (FAFA1), following the basic principles of the FELIX method, and presenting a new optimized implementation algorithm (FAFA2) to improve the circuit-level evaluation criteria significantly was the main novelty of the authors in this paper. We calculate circuit evaluation metrics improvements of the FAFA1 and FAFA2 compared to the exact FELIX-based full adder and different exact and approximate IMPLY-based full adders. In order to ensure the output's accuracy, error analysis metrics in different scenarios are assessed. Then, we apply the FAFA circuit in image processing and average pooling applications and evaluate it according to valid measurements to analyze the quality of images.

The rest of the paper is organized as follows. Section \ref{sec2} will describe the background of memristors, memristive design methods, especially the FELIX one, and also provide an overview of memristor-based adder circuits. The proposed circuit is described in Section \ref{sec3}. Simulations and evaluations of the FAFA1 and FAFA2 are presented in Section \ref{sec4}, and Section \ref{sec5} concludes the paper.

\section{Background} \label{sec2}
\subsection{Memristors} \label{sec21}
\begin{figure}[t]
	\centering
	\includegraphics[scale=0.13]{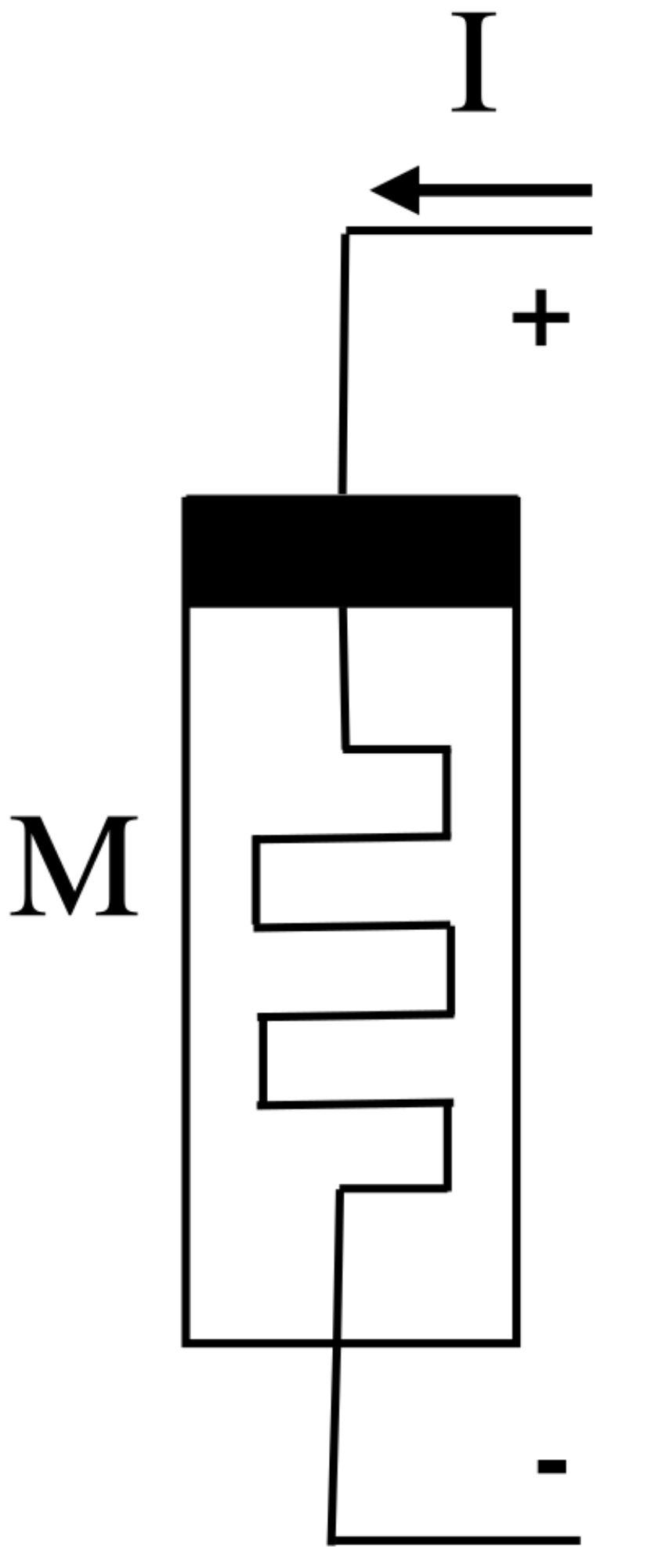}
	\caption{Memristor symbol \cite{ref25}.}
	\label{fig1}
\end{figure}

In 1971, Leon Chua proposed the memristor as the fourth fundamental component linking flux and charge \cite{ref23}. It wasn't until 2008 that Stanley Williams' team at HP Labs successfully constructed the first memristor \cite{ref24}. Following this, researchers have investigated and utilized memristors in various fields. A memristor is a switchable resistor with two electrodes and an insulating active layer between them. Figure \ref{fig1} shows a memristor \cite{ref25}. The resistance of a memristor is influenced by the previous voltage or current that has been applied to it. The value of this resistor can change between two values, $R_{on}$ (low resistance state, logic `1') and $R_{off}$ (high resistance state, logic `0'), based on the voltage or current applied to it. Suppose the current is applied to the Top Electrode (TE) of the memristor (as shown in Figure \ref{fig1}) or the voltage applied to the memristor is greater than the threshold voltage of the memristor. In that case, the memristor's resistance decreases and reaches $R_{on}$. On the other hand, when the current is applied to the Bottom Electrode (BE) of the memristor or a voltage below the negative threshold is applied, the resistance increases, causing the memristor to switch to its $R_{off}$ state.

Among the most attractive memristive design methods, IMPLY \cite{ref26}, FELIX \cite{ref16}, MAGIC \cite{ref27}, Memristor Ratioed Logic (MRL) \cite{ref28}, and reconfigurable NAND/AND \cite{ref29} can be mentioned. IMPLY, FELIX, reconfigurable NAND/AND, and MAGIC logic are categorized in the stateful logic family, while MRL is placed in the non-stateful logic family. In stateful methods, only resistance is used as a state variable, while in non-stateful methods, voltage is also used as a state variable. Stateful design methods are compatible with crossbar array structures. Crossbar arrays are made up of two perpendicular sets of nanowires, with each junction in the structure capable of housing a diode, transistor, resistor, or memristor \cite{ref14}. In this paper, the memristor is placed in the crossbar junctions. Memristor crossbar arrays have low access power consumption \cite{ref10}.

\subsection{FELIX Design Method} \label{sec22}
\begin{figure}[h]
	\centering
	\includegraphics[scale=0.24]{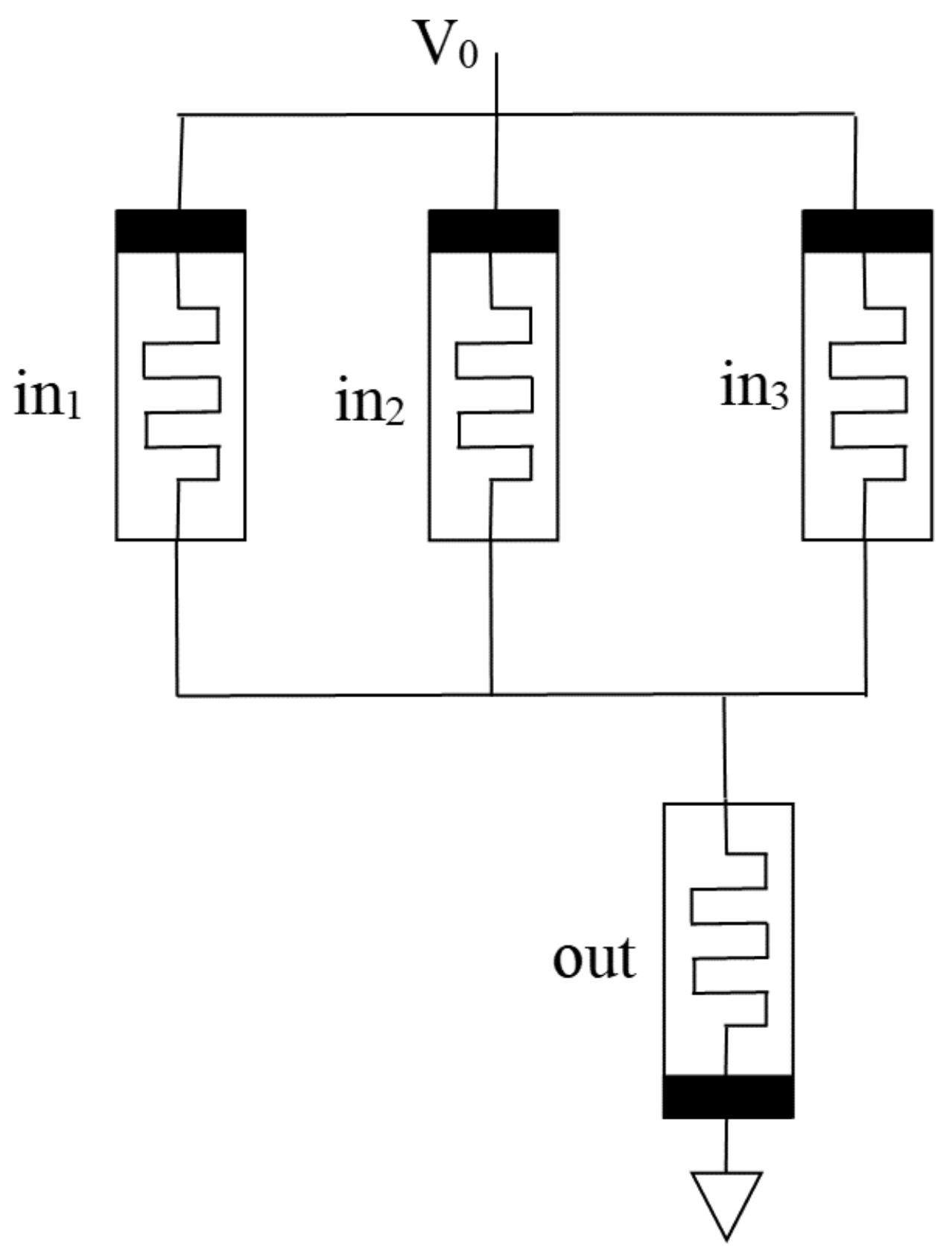}
	\caption{Implementation of 3-input NOR, 3-input NAND, and MIN gates in the FELIX design method \cite{ref16}.}
	\label{fig2}
\end{figure}

Among the memristor-based design methods, one of the inputs is rewritten in the IMPLY method. In the MAGIC method, only the NOR function is implemented in the crossbar array, and all other functions are obtained using NOR, which leads to unnecessary delay overhead. The MRL method is unsuitable for IMC; the FELIX design method is applied in this paper. The FELIX design method enables the activation of IMC and many Boolean functions such as OR, NOT, NOR, Minority (MIN), and NAND in only one cycle and the Majority (MAJ), XOR, and AND functions in two cycles without any additional memristor \cite{ref16}. While implementing OR in MAGIC requires two NOR cycles, and implementing MAJ and AND requires four cycles in this design method \cite{ref16}. Like the MAGIC design method, the FELIX design method preserves inputs, and thus, the input data set can be reused for multiple operations \cite{ref30}. In the FELIX design method, to switch the memristor from $R_{off}$ to $R_{on}$, $V_{pn}>|v_{on}|$ must be established, and $V_{np}>v_{off}$ must be established for switching from $R_{on}$ to $R_{off}$. $|v_{on}|$ and $v_{off}$ are the memristor's threshold voltages, and $V_{pn}$ ($V_{np}$) is the voltage difference between the $p$ ($n$) and $n$ ($p$) terminals \cite{ref16}. In FELIX, the execution voltage, $V_{0}$, is a key factor. A specific $V_{0}$ voltage for each logic function should be applied to operate. In other words, the applied voltage determines the type of operation. Figure \ref{fig2} shows the implementation of 3-input NOR, 3-input NAND, and MIN gates \cite{ref16}. As shown in Figure \ref{fig2}, these three gates have the same implementation, and the applied voltage value, $V_{0}$, determines the type of operation. These operations require only one cycle. For example, implementing a 3-input NOR gate in a memristor-based crossbar array is described. In the first step, the output memristor (labeled out in Figure \ref{fig2}) is initialized to $R_{on}$. In the next step, $V_{0}$ is applied to the TE (terminal $p$) of the inputs, and the TE of the output memristor is connected to the ground.

The value of the execution voltage depends on the model and characteristics of the applied memristor. For example, in \cite{ref16}, 
the execution voltage for 3-input NOR, MIN, and NAND operations is 1V, 0.75V, and 0.67V, respectively, and the value of $R_{on}$ is 10$K\Omega$, and the value of $R_{off}$ is considered equal to 10$M\Omega$. When the applied voltage ($V_{0}$) is set to 1V and $v_{off}$ is 0.5V, the output memristor registers voltages of 0V, 0.5V, 0.67V, and 0.75V for the respective inputs ``000", ``001", ``011", and ``111". This causes the output memristor to maintain its initial value ($R_{on}$) in the first state and switch to $R_{off}$ (logic `0’) in the other seven states, equivalent to a 3-input NOR gate.

\subsection{Memristor-Based Adders} \label{sec23}
Several methods have been proposed in the literature for designing and implementing full adder cells using memristors. These methods include IMPLY, MAGIC, FELIX, reconfigurable NAND/AND, and MRL. The authors in \cite{ref21,ref32,rref1,rref2} introduced IMPLY-based exact full adders using serial architecture. Considering the initialization of work memristors, the full adder in \cite{ref21} needs five memristors. It computes the outputs in 22 computational steps, and the full adder proposed in \cite{ref32} needs five memristors and 23 computational steps to compute $Sum$ and $C_{out}$. The implementation algorithms for both exact full adders proposed in \cite{rref1,rref2} use three work memristors. Due to this feature and the initialization of three work memristors in a single computational step, which differs from the method used in \cite{ref21,ref32}, the outputs of the full adder cells are obtained in 18 \cite{rref1} and 20 \cite{rref2} computational steps and stored in the input memristors. In \cite{ref33}, an exact Carry Select Adder (CSA) and an exact Conditional Carry Adder (CCA) are introduced using parallel architecture and with the IMPLY method. The proposed 4-bit CSA and CCA require 44 and 49 memristors, respectively, and take 39 and 41 computational steps, respectively \cite{ref33}. Compared to serial architecture, parallel architecture has the advantage of being faster. Still, it has disadvantages, such as the need for switches, more memristors, high hardware complexity, and a lack of full compatibility with crossbar arrays \cite{ref1}.

In \cite{ref16}, in addition to introducing the FELIX method and implementing basic logic gates, an exact full adder cell for IMC using XOR and MAJ functions is proposed. In \cite{ref16}, two XOR functions are required to generate $Sum$ output, which takes four cycles and requires two additional memristors. The $C_{out}$ output is also generated after two cycles and requires two additional memristors. It should be mentioned that the initialization of memristors also requires two computational cycles, so the number of cycles increases to eight cycles.

\begin{figure}[h]
	\centering
	\includegraphics[scale=0.24]{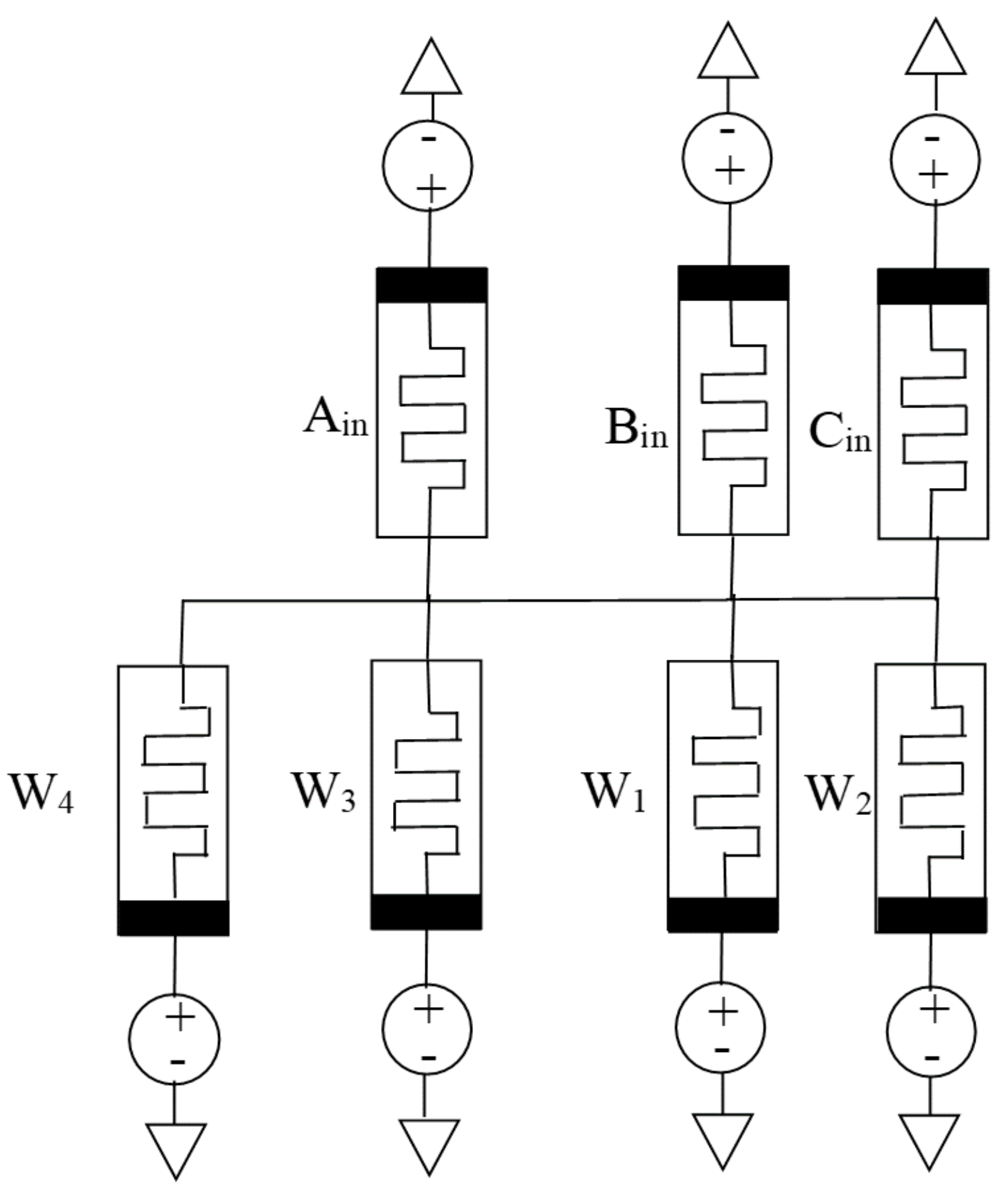}
	\caption{Circuit-level design of the FELIX-based full adder cell in \cite{ref16}.}
	\label{fig3}
\end{figure}

Implementing a full adder using the MAGIC method requires twelve NOR cycles and twelve additional memristors compared to six cycles and four additional memristors in FELIX \cite{ref34}. The operations related to the exact full adder proposed in \cite{ref16} and its memristive circuit are illustrated in Table \ref{tab1} and Figure \ref{fig3}, respectively. In Table \ref{tab1} and Figure \ref{fig3}, $A_{in}$, $B_{in}$, and $C_{in}$ are the inputs of the full adder, and $W_{1}$ to $W_{4}$ are the four additional memristors needed to perform the addition operation.

\begin{table}[h]
	\centering
	\caption{The implementation steps of the exact full adder proposed in \cite{ref16}.}
	\scalebox{1}{
	\begin{tabular}{|c|c|}
		\hline
		Operation & No. of Cycles \\ \hline
		$W_{1}=XOR(A_{in}, B_{in})$ & 2 \\ \hline
		$Sum: W_{2}= XOR(C_{in}, W_{1})$ & 2 \\ \hline
		$W_{3}=MIN(A_{in}, B_{in}, C_{in})$ & 1 \\ \hline
		$C_{out}: W_{4}=NOT(W_{3})$ & 1 \\ \hline 
	\end{tabular}}
	\label{tab1}
\end{table}

The authors have designed a non-stateful approximate full adder using the MRL technique \cite{ref35}. In this memristive approximate full adder, the logical minimization method, which randomly introduces errors in the $Sum$ and $C_{out}$ outputs of the truth table of the exact full adder, is applied to reduce hardware complexity. This proposed design has used ten memristors and two transistors. The authors have stated that buffers were used at the required places due to the voltage drop in the MRL method, which is one of the disadvantages of this method \cite{ref35}. The number of required buffers was not mentioned in \cite{ref35}.

The authors in \cite{ref1,ref22,rref3,rref4} proposed approximate full adders in the serial architecture using the IMPLY method. The design proposed in \cite{ref1} requires four memristors and seven computational steps to calculate the outputs, while the best approximate full adder proposed in \cite{ref22} needs four memristors and eight computational steps. Two serial approximate full adders, SAID1 and SAID2, are introduced using the IMPLY method in \cite{rref4}. Both of these circuits calculate the outputs of the full adder approximately. The computational accuracy of SAID2 is higher than that of the other circuit designed in \cite{rref4}, but the goal of designing SAID1 was to minimize the circuit complexity. The $Sum$ of the SAID1 is obtained in the second computational step without any work memristor, and its $C_{out}$ is also treated as equivalent to the second input of the full adder, without any calculation. The number of computational steps and the number of memristors required for SAID2 are 6 and 5, respectively \cite{rref4}. Several approximate adders are presented in \cite{rref3} using two approximate full adder implementation methodologies. The designs based on the serial architecture that best harmonize with a crossbar array structure are called SINC and SINC+ in \cite{rref3}. The SINC and SINC+ full adders compute the outputs in 3 and 6 steps, respectively, using 3 and 4 memristors. In the four serial full adders, SAID1, SAID2, SINC, and SINC+, the input memristors are overwritten, and if the input data is needed for other computations, a copy operation must be performed before the algorithm is fully executed.

Table \ref{tab2} summarizes the full adders mentioned. The ED is the difference between the exact and approximate values, the ER is the ratio of the number of states in which the output is incorrect to the total possible states, the MED is the mean ED of all inputs, and the NMED is the normalized value of MED \cite{ref22,ref36}.

\begin{table}[!]
	\centering
	\caption{Summary of mentioned approximate and exact full adders.}
	\scalebox{0.9}{
	\begin{tabular}{|c|c|c|c|c|c|c|c|c|c|c|}
		\hline
		Full & ED & $ER_{Sum}$ & $ER_{Cout}$ & MED & NMED & No. of & No. of & No. of & Stateful & Mode \\
		Adder & & & & & & Memristors & Steps & Resistors & &\\ \hline
		\cite{ref1} & 3 & $\frac{3}{8}$ & $\frac{1}{8}$ & 0.375 & 0.125 & 4 & 7 & 1 & \ding{51} (IMPLY) & Approximate \\ \hline
		\cite{ref22} & 3 & $\frac{3}{8}$ & $\frac{1}{8}$ & 0.375 & 0.125 & 4 & 8 & 1 & \ding{51} (IMPLY) & Approximate \\ \hline		
		SAID1 \cite{rref4} & 4 & $\frac{4}{8}$ & $\frac{2}{8}$ & 0.5 & 0.166 & 3 & 2 & 1 & \ding{51} (IMPLY) & Approximate \\ \hline
		SAID2 \cite{rref4} & 3 & $\frac{3}{8}$ & $\frac{2}{8}$ & 0.375 & 0.125 & 5 & 6 & 1 & \ding{51} (IMPLY) & Approximate \\ \hline	
		SINC \cite{rref3} & 6 & $\frac{4}{8}$ & $\frac{4}{8}$ & 0.75 & 0.25 & 3 & 3 & 1 & \ding{51} (IMPLY) & Approximate \\ \hline
		SINC+ \cite{rref3} & 4 & $\frac{4}{8}$ & $\frac{2}{8}$ & 0.5 & 0.166 & 4 & 6 & 1 & \ding{51} (IMPLY) & Approximate \\ \hline		
		\cite{ref35} & 7 & $\frac{3}{8}$ & $\frac{2}{8}$ & 0.875 & 0.291 & 10 & - & - & \ding{53} (MRL) & Approximate \\ \hline
		\cite{ref21} & 0 & 0 & 0 & 0 & 0 & 5 & 22 & 1 & \ding{51} (IMPLY) & Exact \\ \hline
		\cite{ref32} & 0 & 0 & 0 & 0 & 0 & 5 & 23 & 1 & \ding{51} (IMPLY) & Exact \\ \hline
		\cite{rref1} & 0 & 0 & 0 & 0 & 0 & 6 & 18 & 1 & \ding{51} (IMPLY) & Exact \\ \hline
		\cite{rref2} & 0 & 0 & 0 & 0 & 0 & 6 & 20 & 1 & \ding{51} (IMPLY) & Exact \\ \hline
		\cite{ref16} & 0 & 0 & 0 & 0 & 0 & 7 & 8 & - & \ding{51} (FELIX) & Exact \\ \hline
		\cite{ref34} & 0 & 0 & 0 & 0 & 0 & 15 & 13 & - & \ding{51} (MAGIC) & Exact \\ \hline
	\end{tabular}}
	\label{tab2}
\end{table}

\section{Proposed Memristive Approximate Full Adder Based On FELIX} \label{sec3}
The full adder circuit takes three inputs ($A_{in}$, $B_{in}$, and $C_{in}$) and produces two outputs, $Sum$ and $C_{out}$. $Sum$ and $C_{out}$ outputs of the exact full adder circuit can be computed using Eqs. (\ref{eq1}) and (\ref{eq2}), respectively. These equations are computed based on the truth table of the exact circuit.

\begin{gather}
	Sum=\overline{A_{in}} \cdot \overline{B_{in}} \cdot C_{in} +  A_{in} \cdot \overline{B_{in}} \cdot \overline{C_{in}} + \overline{A_{in}} \cdot B_{in} \cdot \overline{C_{in}} + A_{in} \cdot B_{in} \cdot C_{in} = A_{in} \oplus B_{in} \oplus C_{in} \label{eq1} \\
	C_{out}= A_{in} \cdot B_{in} + A_{in} \cdot C_{in} + B_{in} \cdot C_{in} = A_{in} \cdot B_{in} + C_{in} \cdot (A_{in} \oplus B_{in}) \label{eq2}
\end{gather}

In this paper, the FAFA is proposed in order to decrease the hardware complexity and the delay of the memristive FELIX-based exact full adder circuit for IMC. The FAFA is designed by converting the two rows of the exact full adder’s truth table. The truth table of the FAFA is shown in Table \ref{tab3}. Exact states are represented by \ding{51}, and \ding{53} represents inexact states. In the FAFA, the $C_{out}$ output is error-free (equal to the exact $C_{out}$ output), and the ER of the FAFA’s $Sum$ is 0.25. Also, the ED, MED, and NMED of the FAFA are 2, 0.25, and 0.083, respectively. The FAFA is implemented using the FELIX design method compatible with the crossbar array. According to the truth table of the FAFA, the $Sum$ output is equivalent to the MIN function, and the $C_{out}$ output is equivalent to the MAJ function. Therefore, after obtaining the $Sum$, the $C_{out}$ output can be easily obtained by complementing it. The equations related to the outputs are shown in Eqs. (\ref{eq3}) and (\ref{eq4}).

\begin{gather}
	Sum=MIN(A_{in}, B_{in}, C_{in}) \label{eq3} \\
	C_{out}= MAJ(A_{in}, B_{in}, C_{in})=NAND(Sum, 1)=NOT(Sum)=\overline{Sum} \label{eq4}
\end{gather}

\begin{table}[!]
	\centering
	\caption{The truth table of the FAFA.}
	\scalebox{1}{
	\begin{tabular}{|c|c|c|c|c|c|c|}
		\hline
		$A_{in}$ & $B_{in}$ & $C_{in}$ & Exact $Sum$ & Exact $C_{out}$ & Proposed $Sum$ & Proposed $C_{out}$ \\ \hline
		0 & 0 & 0 & 0 & 0 & 1 \ding{53} & 0 \ding{51} \\ \hline
		0 & 0 & 1 & 1 & 0 & 1 \ding{51} & 0 \ding{51} \\ \hline
		0 & 1 & 0 & 1 & 0 & 1 \ding{51} & 0 \ding{51} \\ \hline
		0 & 1 & 1 & 0 & 1 & 0 \ding{51} & 1 \ding{51} \\ \hline
		1 & 0 & 0 & 1 & 0 & 1 \ding{51} & 0 \ding{51} \\ \hline
		1 & 0 & 1 & 0 & 1 & 0 \ding{51} & 1 \ding{51} \\ \hline
		1 & 1 & 0 & 0 & 1 & 0 \ding{51} & 1 \ding{51} \\ \hline
		1 & 1 & 1 & 1 & 1 & 0 \ding{53} & 1 \ding{51} \\ \hline
	\end{tabular}}
	\label{tab3}
\end{table}

Two approaches have been chosen to implement the proposed circuit. In the FAFA1, the proposed circuit requires a total of six memristors (three input memristors, $A_{in}$, $B_{in}$, and $C_{in}$, and three additional memristors, $W_{1}$, $W_{2}$, and $W_{3}$), and the whole operation is computed in two cycles. One cycle is required to calculate the MIN function, and one must perform the NAND(MIN, `1') operation. This approach requires an additional memristor to perform the NAND operation, which is initialized to logic `1' ($R_{on}$). 

\begin{figure}[!]
     \centering
     \begin{subfigure}[]{0.375\textwidth}
         \centering
         \includegraphics[width=0.7\textwidth]{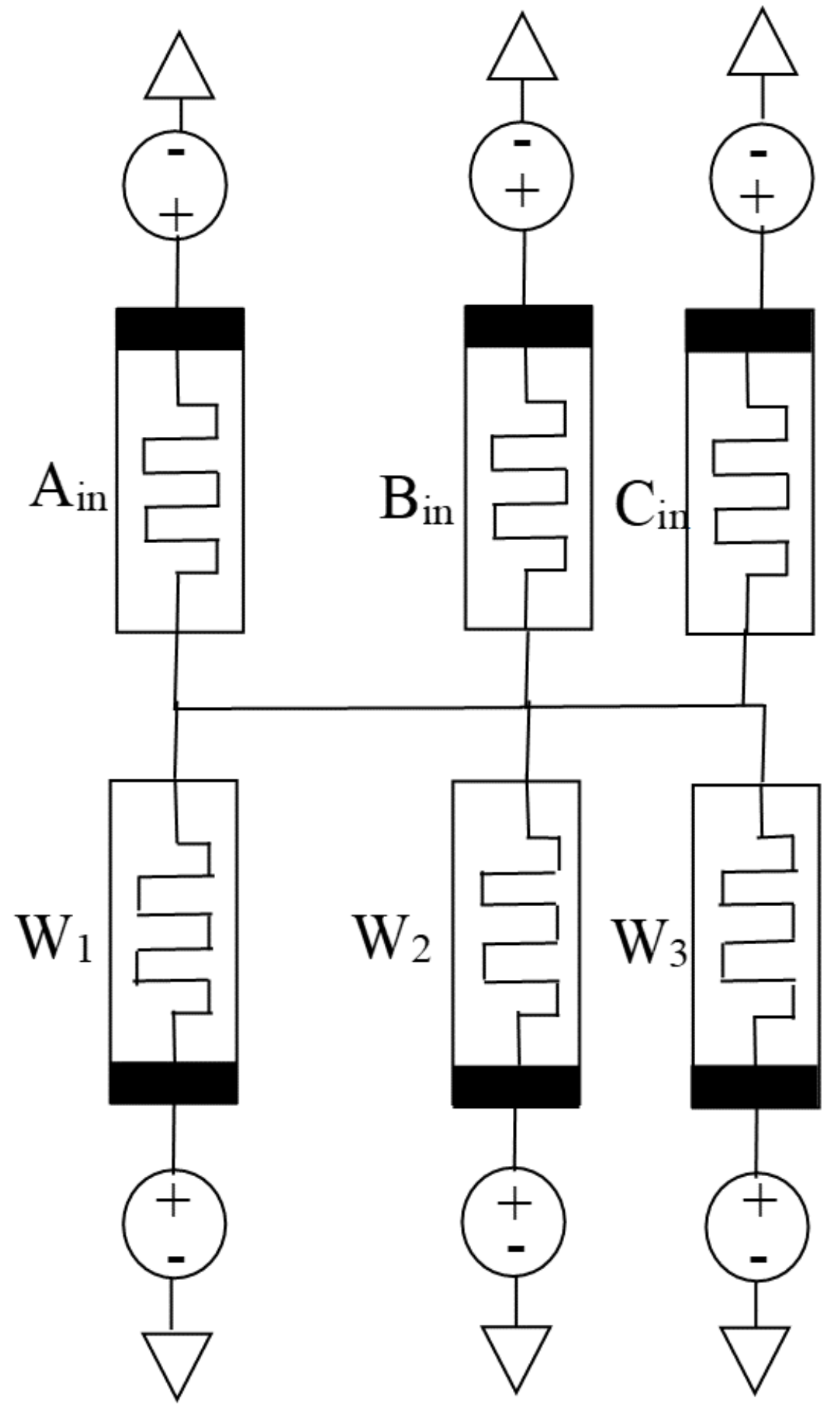}
         \caption{}
         \label{Fig4a}
     \end{subfigure}
     \begin{subfigure}[]{0.35\textwidth}
         \centering
         \includegraphics[width=0.7\textwidth]{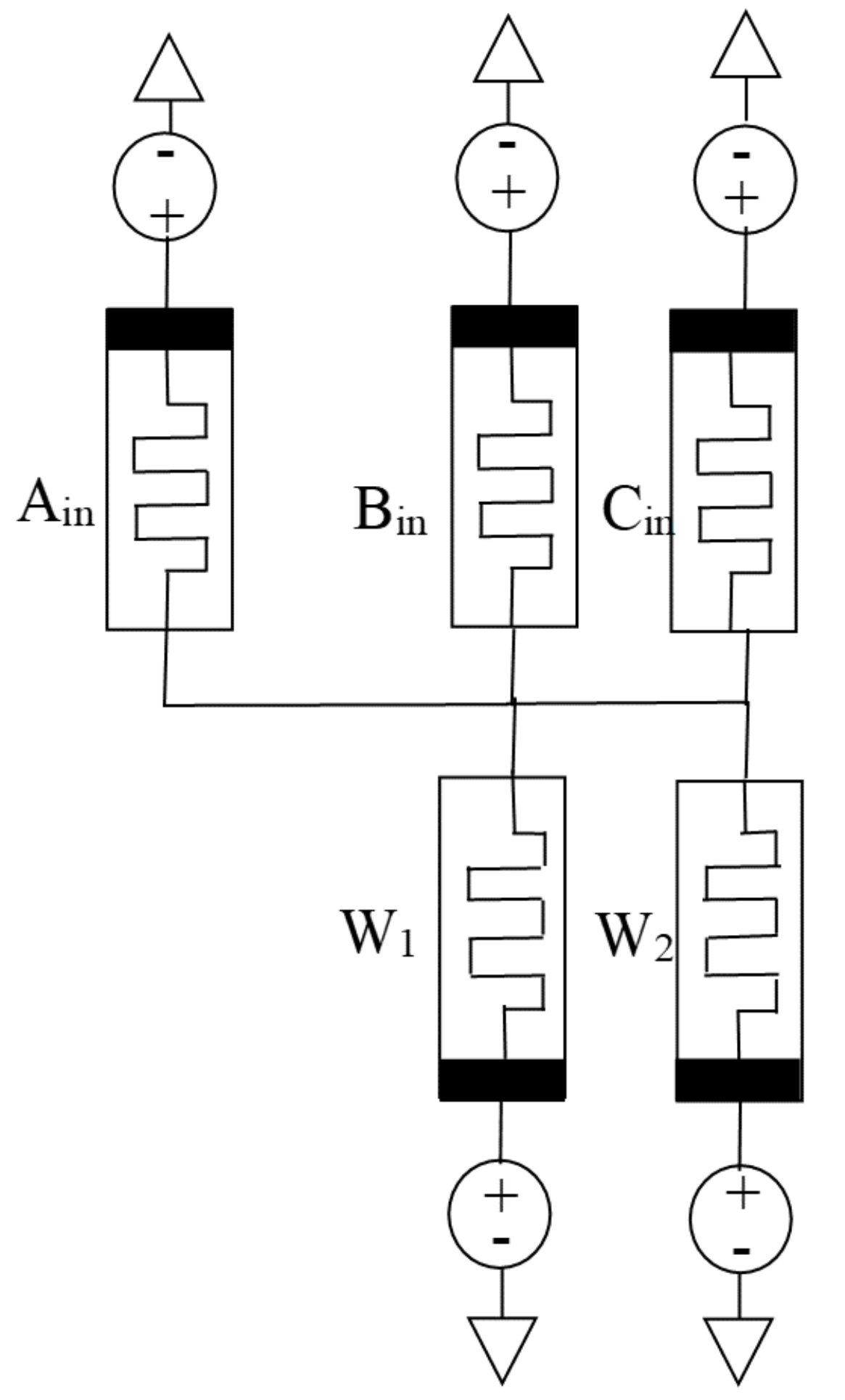}
         \caption{}
         \label{Fig4b}
     \end{subfigure}
        \caption{Circuit-level design of the FAFA: (a) the FAFA1, and (b) the FAFA2.}
        \label{Fig4}
\end{figure}

In the second approach, the proposed circuit involves five memristors (three input memristors, $A_{in}$, $B_{in}$, and $C_{in}$, and two additional memristors, $W_{1}$ and $W_{2}$). The whole operation is performed in two cycles: one to calculate the MIN function and one to perform the NOT operation. In both implementation approaches, outputs are calculated in three cycles if the initialization of memristors is considered. The implementation algorithms of the FAFA1 and FAFA2 are written in Tables \ref{tab4} and \ref{tab5}, respectively, and their circuit-level designs are shown in Figures \ref{Fig4}(a) and \ref{Fig4}(b).

\begin{table}[t]
	\centering
	\caption{The implementation steps of the FAFA1.}
	\scalebox{1}{
	\begin{tabular}{|c|c|c|}
		\hline
		Step & Operation & No. of Required Cycles \\ \hline
		1 & $Sum: W_{1}= MIN(A_{in}, B_{in}, C_{in})$ & 1 \\ \hline
		2 & $C_{out}: W_{2}=NAND(W_{1}, 1)$ & 1 \\ \hline 
	\end{tabular}}
	\label{tab4}
\end{table}

\begin{table}[t]
	\centering
	\caption{The implementation steps of the FAFA2.}
	\scalebox{1}{
	\begin{tabular}{|c|c|c|}
		\hline
		Step & Operation & No. of Required Cycles \\ \hline
		1 & $Sum: W_{1}= MIN(A_{in}, B_{in}, C_{in})$ & 1 \\ \hline
		2 & $C_{out}: W_{2}=NOT(W_{1})$ & 1 \\ \hline 
	\end{tabular}}
	\label{tab5}
\end{table}

\section{Simulations And Evaluations Of The Proposed Circuit} \label{sec4}
\subsection{Circuit-Level Simulation And Analysis} \label{sec41}
The FAFA is simulated in both implementation approaches: in the LTSPICE simulator and using the model introduced in \cite{ref37}. The memristor model proposed in \cite{ref37} is a general voltage-controlled behavioral model. Its mathematical equations are written in Eqs. (\ref{eq5}-\ref{eq7}). In the circuit-level simulations, the values of $a$, $R_{on}$, and $R_{off}$ equal 10, 100$\Omega$, and 1000$\Omega$, respectively.

\begin{equation}
\frac{dx}{dt} =
\begin{cases} 
    \frac{1}{e^{x-a}}, & \text{if } v > 1.2 \\
    0, & \text{if } -0.6 \leq v \leq 1.2 \\
    \frac{-1}{e^{-x-a}}, & \text{if } v < -0.6
\end{cases}
\label{eq5}
\end{equation}

\begin{equation}
R(x) =
\begin{cases} 
    R_{on}, & \text{if } x \geq 0 \\
    R_{off}, & \text{if } x < 0
\end{cases}
\label{eq6}
\end{equation}

\begin{equation}
i(t) = \frac{1}{R(x)}v(t)
\label{eq7}
\end{equation}

In the behavioral model applied in this paper to simulate the proposed cell (in both approaches), changing the values of $R_{on}$, $R_{off}$, and parameter $a$ directly affects the performance of the circuits. The circuit response and switching of the applied memristors strongly depend on the threshold voltages \cite{ref37}. As $x$ is applied to control the resistance ($R(x)$) and a logarithmic function defines the connection between $x$ and $t$, the value of $x$ has rapid changes \cite{ref37}. Also, the conversion time of the memristor's resistance between $R_{on}$ and $R_{off}$ depends on $a$ \cite{ref37}. In this paper, the parameters reported in \cite{ref37}, which were used to simulate the MAGIC-based basic logic gates, have been used to simulate the FELIX-based exact \cite{ref16} and approximate circuits. The authors' goal in choosing these parameters was to ensure the appropriate response of the circuits in different input states and optimal curve fitting \cite{ref37}. The FELIX method is stateful, in which the determination of threshold voltages is affected by the logic function and the number of inputs \cite{ref16}. Therefore, changing the memristor model can affect the functionality of the evaluated FELIX-based circuits. Investigating the effect of process variation and assessing the applied memristor's retention time on the performance and behavior of the proposed FELIX-based cells is one of the technical aspects in evaluating the proposed circuits, which is considered a valuable future work considering the context of the article and the perspective of presenting the circuits' implementation algorithm and its verification.

\begin{figure}[t]
    \centering

    \begin{subfigure}{0.48\textwidth}
        \centering
        \includegraphics[width=\linewidth]{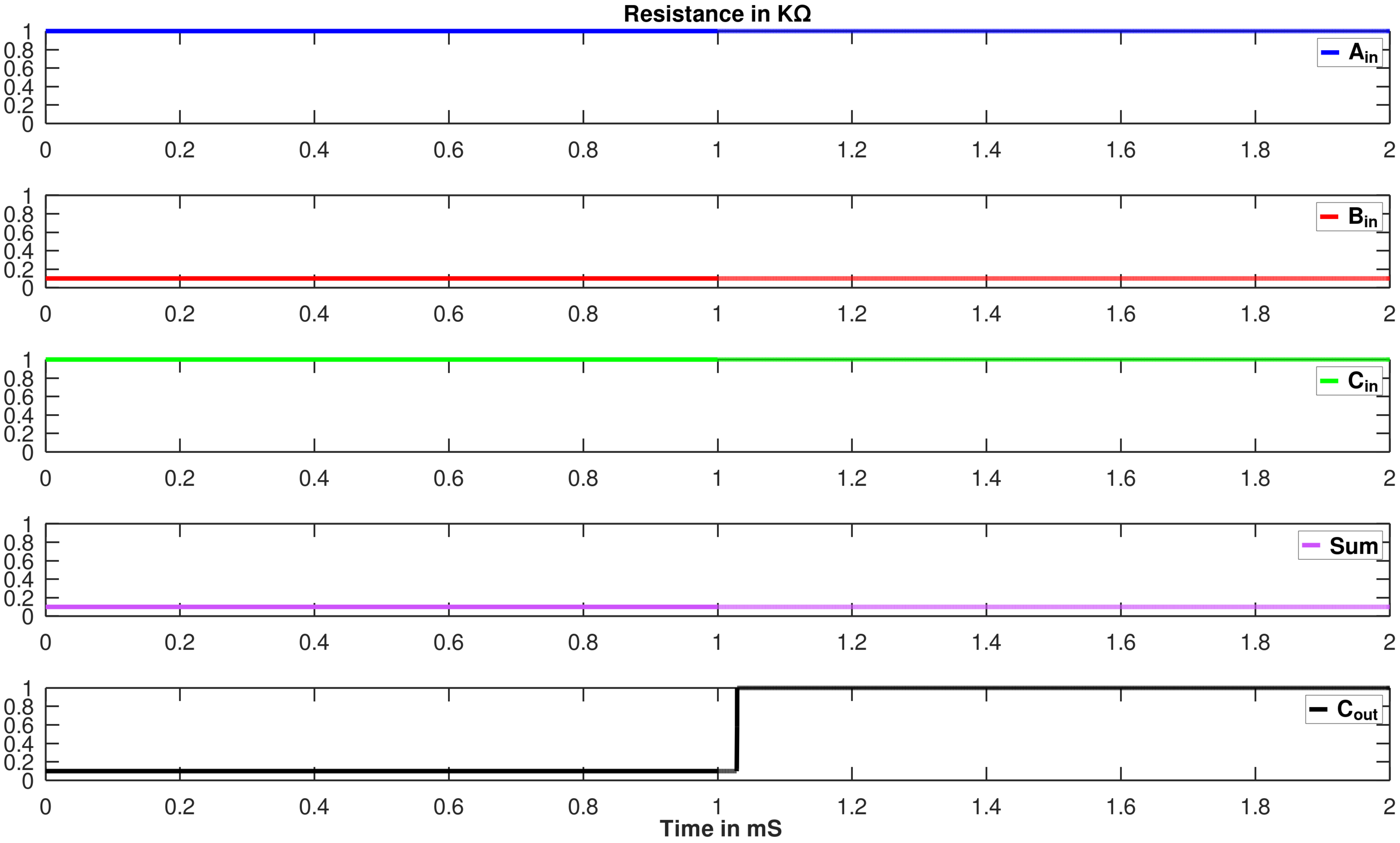}
        \caption{}
    \end{subfigure}
    \hfill
    \begin{subfigure}{0.48\textwidth}
        \centering
        \includegraphics[width=\linewidth]{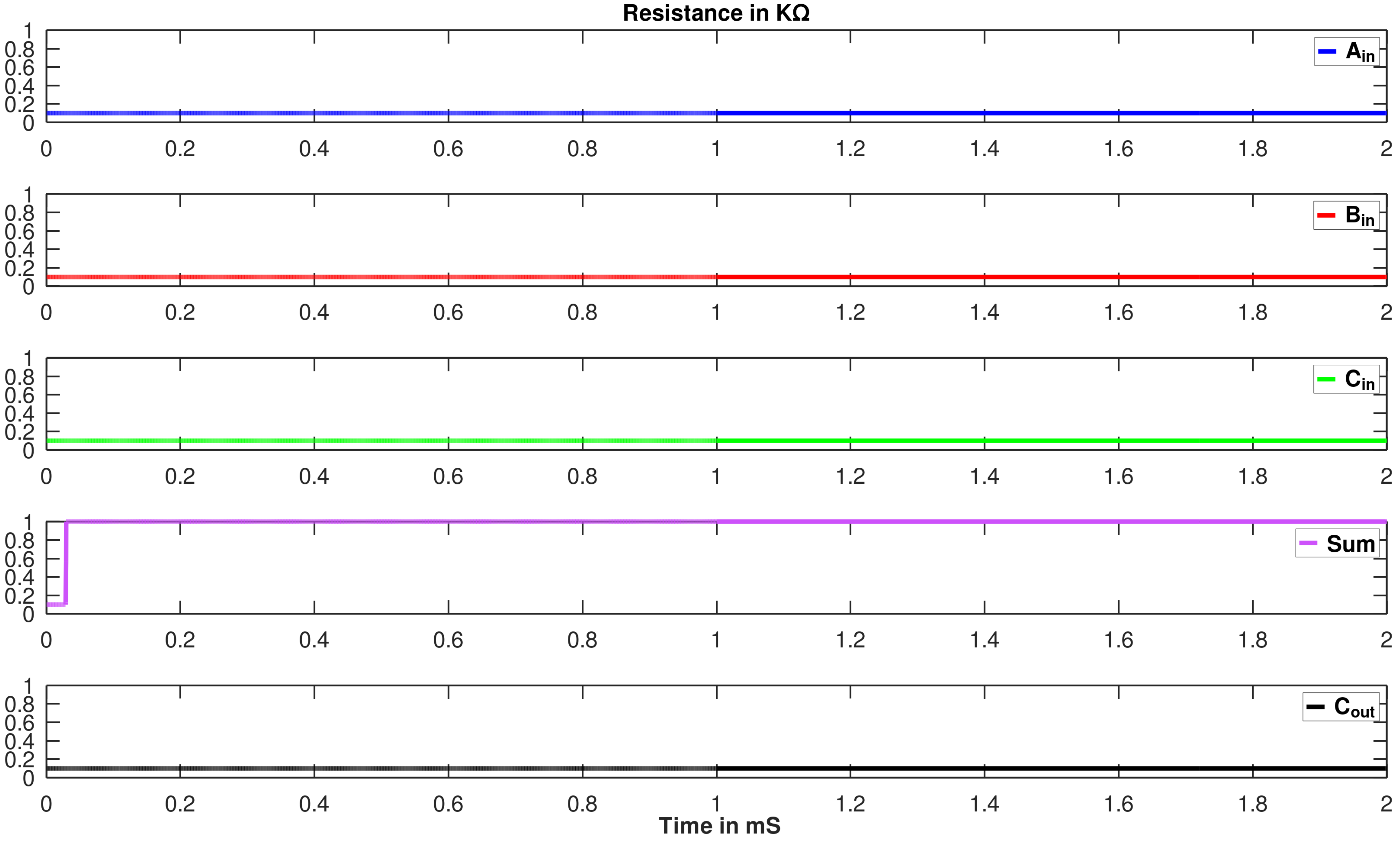}
        \caption{}
    \end{subfigure}

    \begin{subfigure}{0.48\textwidth}
        \centering
        \includegraphics[width=\linewidth]{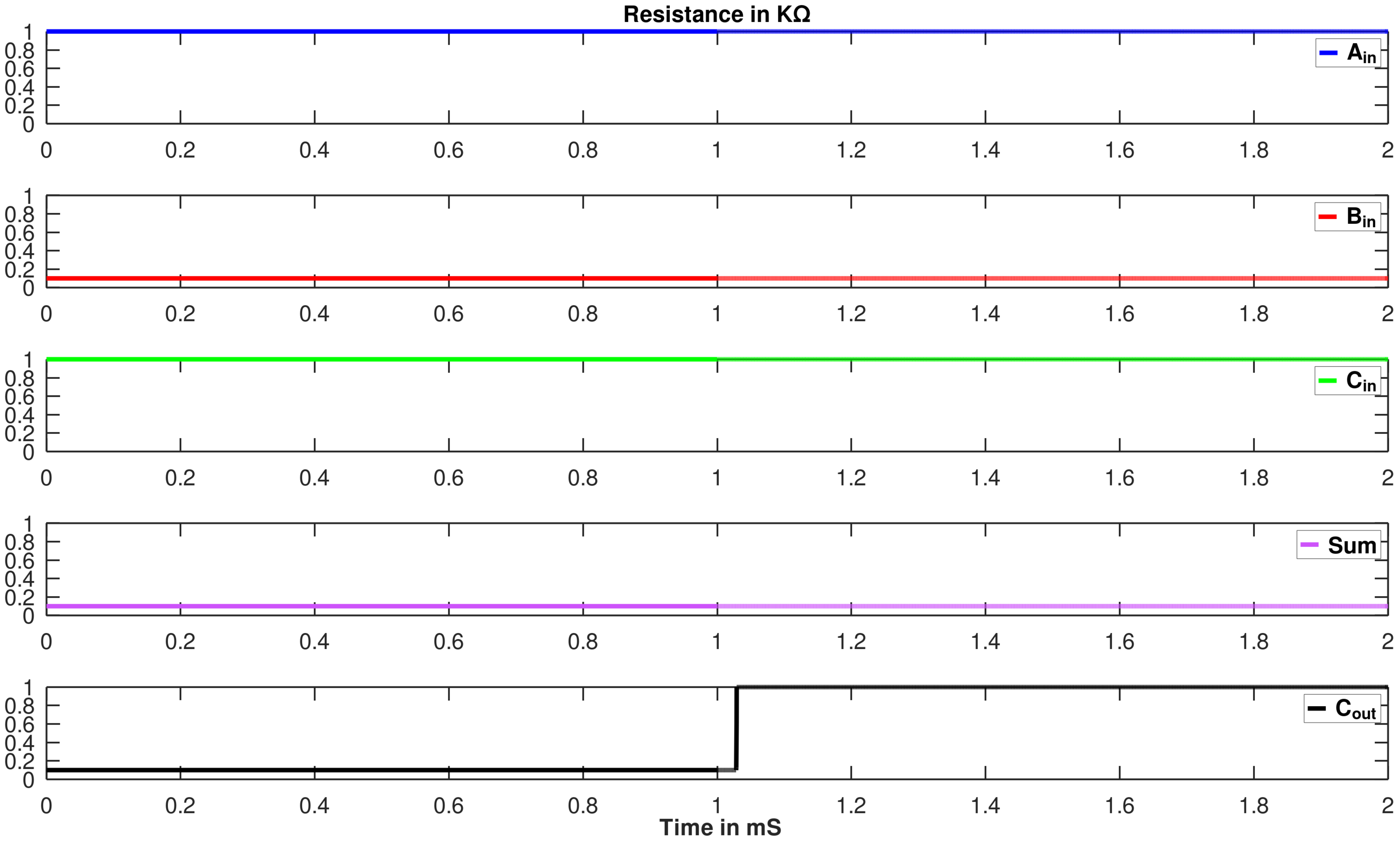}
        \caption{}
    \end{subfigure}
    \hfill
    \begin{subfigure}{0.48\textwidth}
        \centering
        \includegraphics[width=\linewidth]{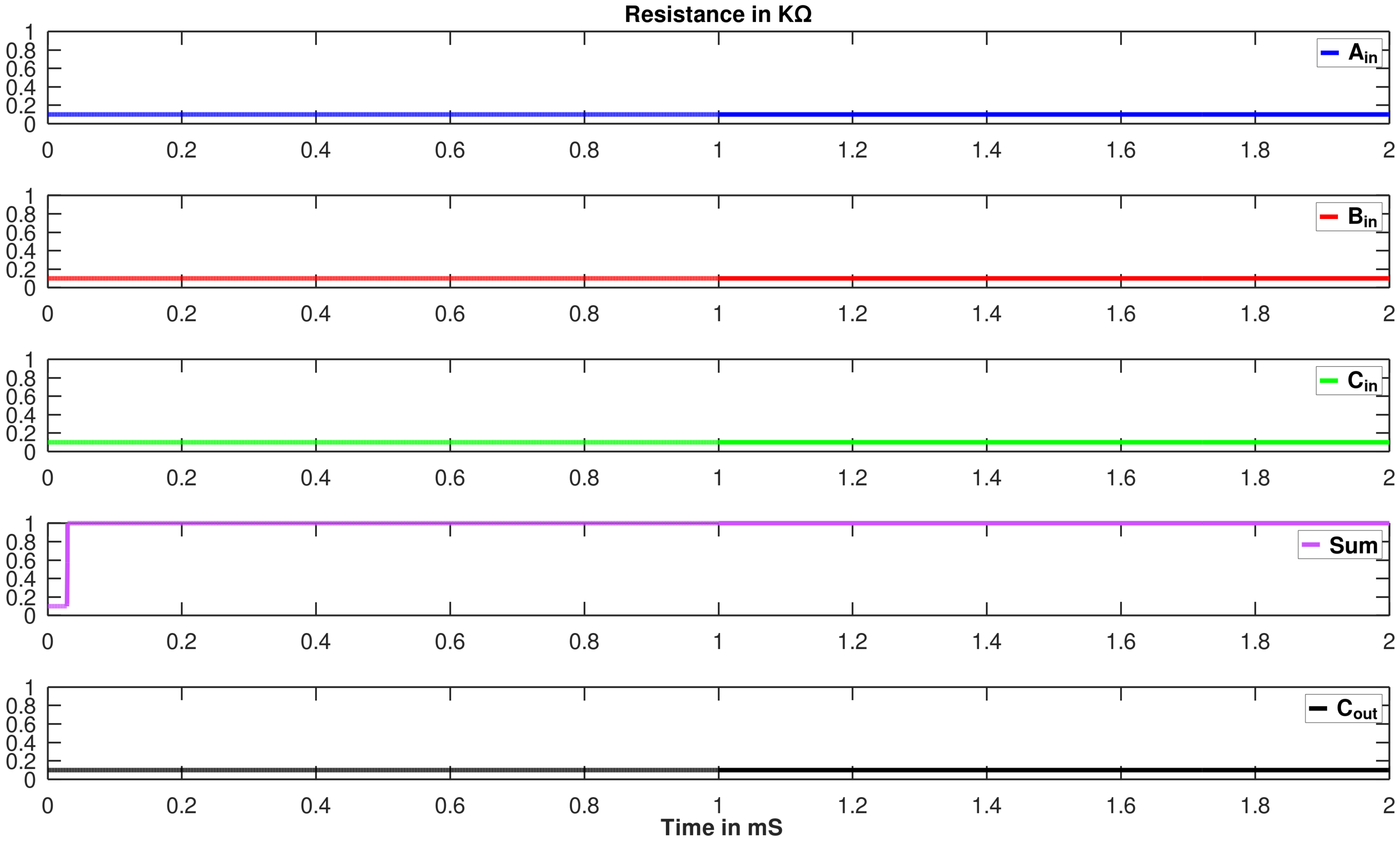}
        \caption{}
    \end{subfigure}

        \caption{The output waveforms of the proposed full adders: (a) FAFA1: $A_{in}B_{in}C_{in} =``010"$, (b) FAFA1: $A_{in}B_{in}C_{in}=``111"$, (c) FAFA2: $A_{in}B_{in}C_{in} =``010"$, and (d) FAFA2: $A_{in}B_{in}C_{in}=``111".$}
        \label{Fig5}
\end{figure}

The LTSPICE simulator verified the correctness of the outputs of all eight possible input states of the FAFA1 and FAFA2. Circuit-level simulation results showed that the proposed algorithm produces outputs corresponding to the proposed truth table in all states. The outputs of the proposed circuit in two input states, $A_{in}B_{in}C_{in}$=``111" and $A_{in}B_{in}C_{in}$=``010", in the first and second approaches are shown in subfigures \ref{Fig5}(a)-\ref{Fig5}(d).

The values of the execution voltages ($V_{0}$) applied in the implementation of the FAFA1 and FAFA2 are shown in Table \ref{tab6}. The FAFA is compared with the exact one \cite{ref16} regarding the memristors` count, cycles, and energy consumption in both implementation approaches. The comparison is tabulated in Table \ref{tab7}. The implementation of FAFA1 and FAFA2 has reduced the number of required memristors by 14.28\% and 28.57\%, respectively, compared to the full adder mentioned in \cite{ref16}. The number of cycles required to implement the FAFA1 and FAFA2 has improved by 66.66\% compared to the adder proposed in \cite{ref16}. The method introduced in \cite{ref1,ref22,ref38,nref1,nref3,nref4} is applied to calculate the energy consumption. First, the total energy consumption of the memristors involved in the operation in all eight states of the truth table in each input state has been calculated. Then, the average energy consumption value of all input states is reported as the energy consumption of each circuit. The energy consumption of the FAFA1 and FAFA2 has been reduced by 73.735\% and 81.754\%, respectively, compared to the full adder proposed in \cite{ref16}. 

\begin{table}[b]
	\centering
	\caption{The execution voltage ($V_{0}$) of basic functions applied in the implementation of the FAFA1 and FAFA2.}
	\scalebox{1}{
	\begin{tabular}{|c|c|c|c|}
		\hline
		Gate & MIN & NAND & NOT \\ \hline
		The value of $V_{0}$ & 1.2V & 1.2V & 1.55V  \\ \hline
	\end{tabular}}
	\label{tab6}
\end{table}

Table \ref{tab7} lists the circuit characteristics of the stateful memristive exact and approximate full adders proposed in \cite{ref1,ref21,ref22,rref1,rref2,rref3,rref4}, which are implemented using the IMPLY design method in serial architecture. This table compares only the number of required memristors and the computational steps of the proposed approximate full adders (FAFA1 and FAFA2) with the exact and approximate IMPLY-based full adders. It should be noted that the initialization steps are not included in the computational cycles of the exact and approximate full adders reported in Table \ref{tab7}. Due to the difference in the memristor model used in the circuit level simulation of this paper with the proposed circuits in \cite{ref1,ref21,ref22,rref1,rref2,rref3,rref4}, it is not possible to directly compare the energy consumption of these circuits with each other. The number of memristors applied in the IMPLY-based exact and approximate full adders proposed in \cite{ref1,ref21,ref22,rref1,rref2,rref3,rref4} is at most 50\% less than the number of memristors applied in the proposed circuits here. However, it should be noted that the IMPLY method is destructive, and the input values stored in the input memristors are erased after the implementation algorithm is executed. Therefore, if input data needs to be stored, a copy operation (two NOT operations for each input bit) must be performed using work memristors, significantly increasing the number of computational steps and the number of memristors. The FELIX method is non-destructive, and there is no need to perform a copy operation to preserve the data after executing the proposed circuit implementation algorithm. Proposed approximate full adders have improved the number of computational steps by up to 90\% compared to the proposed circuits in \cite{ref1,ref21,ref22,rref1,rref2,rref3,rref4}. The mentioned percentages are reported without considering the computational steps needed to store the input data in the IMPLY-based exact and approximate full adders \cite{ref1,ref21,ref22,rref1,rref2,rref3,rref4}.

\begin{table}[!]
	\centering
	\caption{Exact and approximate full adders' circuit-level criteria comparison.}
	\scalebox{0.85}{
	\begin{tabular}{|c|c|c|c|c|c|c|}
		\hline
		Full Adder & No. Of Memristors & Improvement & No. Of Cycles & Improvement & Energy Consumption & Improvement \\
		& & Over \cite{ref16} & & Over \cite{ref16} & & Over \cite{ref16} \\ \hline
		Exact FELIX \cite{ref16} & 7 & - & 6 & - & 60.679 $\mu$J & - \\ \hline
		Exact IMPLY \cite{ref21} & 5 & 28.57\% & 20 & -233.33\% & NC\textsuperscript{*} & NC\textsuperscript{*} \\ \hline
		Exact IMPLY \cite{rref1} & 6 & 14.28\% & 17 & -183.33\% & NC\textsuperscript{*} & NC\textsuperscript{*} \\ \hline
		Exact IMPLY \cite{rref2} & 6 & 14.28\% & 19 & -216.66\% & NC\textsuperscript{*} & NC\textsuperscript{*} \\ \hline
		SIAFA1 \cite{ref22} & 4 & 42.85\% & 7 & -16.66\% & NC\textsuperscript{*} & NC\textsuperscript{*} \\ \hline
		SIAFA2 \cite{ref22} & 5 & 28.57\% & 8 & -33.33\% & NC\textsuperscript{*} & NC\textsuperscript{*} \\ \hline
		SIAFA3 \cite{ref22} & 4 & 42.85\% & 7 & -16.66\% & NC\textsuperscript{*} & NC\textsuperscript{*} \\ \hline
		SIAFA4 \cite{ref22} & 4 & 42.85\% & 7 & -16.66\% & NC\textsuperscript{*} & NC\textsuperscript{*} \\ \hline
		SAFAN \cite{ref1} & 4 & 42.85\% & 6 & 0\% & NC\textsuperscript{*} & NC\textsuperscript{*} \\ \hline
		SAID1 \cite{rref4} & 3 & 57.14\% & 2\textsuperscript{+} & 66.66\% & NC\textsuperscript{*} & NC\textsuperscript{*} \\ \hline
		SAID2 \cite{rref4} & 5 & 28.57\% & 4 & 33.33\% & NC\textsuperscript{*} & NC\textsuperscript{*} \\ \hline
		SINC \cite{rref3} & 3 & 57.14\% & 2 & 66.66\% & NC\textsuperscript{*} & NC\textsuperscript{*} \\ \hline
		SINC+ \cite{rref3} & 4 & 42.85\% & 5 & 16.66\% & NC\textsuperscript{*} & NC\textsuperscript{*} \\ \hline
		FAFA1 & 6 & 14.28\% & 2 & 66.66\% & 15.937 $\mu$J & 73.735\% \\ \hline
		FAFA2 & 5 & 28.57\% & 2 & 66.66\% & 11.071 $\mu$J & 81.754\% \\ \hline
	\end{tabular}}
	\label{tab7}
	\begin{flushleft} \textsuperscript{*} \footnotesize Not Calculated. \end{flushleft}
	\begin{flushleft} \textsuperscript{+} \footnotesize The initialization step must be considered because one of the input memristors is directly initialized. \end{flushleft}
\end{table}

\begin{figure}[t]
	\centering
	\includegraphics[scale=0.7]{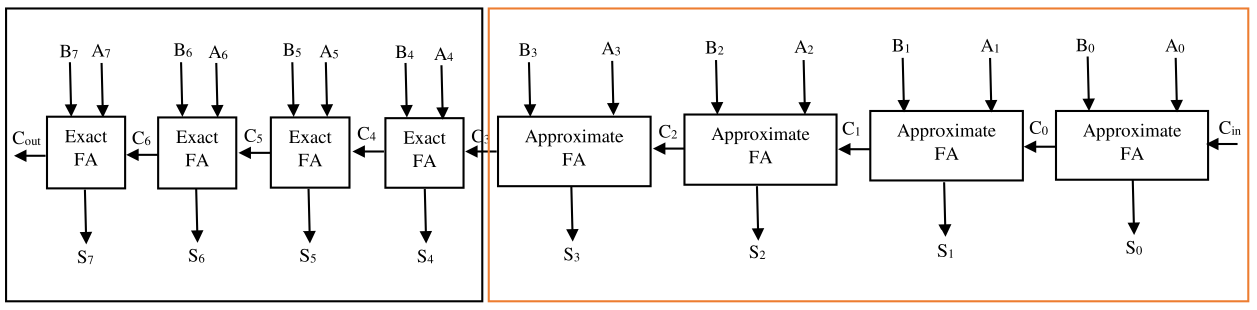}
	\caption{The structure of the $1^{st}$ scenario of the 8-bit approximate RCA, four LSBs are approximate and four MSBs are exact \cite{ref1}.}
	\label{fig7}
\end{figure}

For a comprehensive analysis of the FAFA, the 8-bit approximate Ripple Carry Adder (RCA) has been evaluated in two different scenarios. This approximate adder was divided into two exact and approximate parts in both scenarios. In the exact part of the $1^{st}$ scenario’s structure, the exact full adder was applied to the four MSBs. The FAFA was applied in the LSBs in this scenario. In the $2^{nd}$ scenario, the exact section contains three exact full adders in the MSBs, and the other five are approximate ones (LSBs). In both scenarios, the exact full adder proposed in \cite{ref16} is used for the MSBs. The $1^{st}$ scenario’s structure is drawn in Figure \ref{fig7}. The memristor count, cycles, and energy consumption (excluding the energy required to initialize the memristors) of the approximate 8-bit RCAs (consisting of FAFA1 and FAFA2) in both scenarios are compared with the exact 8-bit RCA built using the exact full adder proposed in \cite{ref16} in Table \ref{tab8}.

\begin{table}[h]
	\centering
	\caption{Comparison of the number of memristors, the number of cycles, and the energy consumption of the 8-bit approximate and exact RCAs with the 8-bit exact RCA proposed in \cite{ref16}.}
	\scalebox{0.9}{
	\begin{tabular}{|c|c|c|c|c|c|c|}
		\hline
		8-bit Serial & No. Of Memristors & Improvement & No. Of Cycles & Improvement & Energy Consumption & Improvement \\
		RCA & & Over \cite{ref16} & & Over \cite{ref16} & & Over \cite{ref16} \\ \hline
		\cite{ref16} & 28 & - & 64 & - & 485.432 $\mu$J & - \\ \hline
		\cite{ref21} & 19 & 32.15\% & 176 & -175\% & NC\textsuperscript{*} & NC\textsuperscript{*} \\ \hline
		\cite{rref1} & 20 & 28.57\% & 144 & -125\% & NC\textsuperscript{*} & NC\textsuperscript{*} \\ \hline
		\cite{rref2} & 20 & 28.57\% & 160 & -150\% & NC\textsuperscript{*} & NC\textsuperscript{*} \\ \hline
		\multicolumn{7}{|c|}{$1^{st}~scenario:~the~four~MSBs~are~exact,~and~the~four~LSBs~are~approximate.$} \\ \hline
		SIAFA1 \cite{ref22} & 19 & 32.14\% & 120 & -87.5\% & NC\textsuperscript{*} & NC\textsuperscript{*} \\ \hline
		SIAFA2 \cite{ref22} & 19 & 32.14\% & 128 & -100\% & NC\textsuperscript{*} & NC\textsuperscript{*} \\ \hline
		SIAFA3 \cite{ref22} & 19 & 32.14\% & 120 & -87.5\% & NC\textsuperscript{*} & NC\textsuperscript{*} \\ \hline
		SIAFA4 \cite{ref22} & 19 & 32.14\% & 120 & -87.5\% & NC\textsuperscript{*} & NC\textsuperscript{*} \\ \hline
		SAFAN \cite{ref1} & 19 & 32.14\% & 116 & -81.25\% & NC\textsuperscript{*} & NC\textsuperscript{*} \\ \hline
		SAID1 \cite{rref4} & 19 & 32.14\% & 96 & -50\% & NC\textsuperscript{*} & NC\textsuperscript{*} \\ \hline
		SAID2 \cite{rref4} & 23 & 17.86\% & 112 & -75\% & NC\textsuperscript{*} & NC\textsuperscript{*} \\ \hline
		SINC \cite{rref3} & 19 & 32.14\% & 100 & -56.25\% & NC\textsuperscript{*} & NC\textsuperscript{*} \\ \hline
		SINC+ \cite{rref3} & 19 & 32.14\% & 103 & -60.94\% & NC\textsuperscript{*} & NC\textsuperscript{*} \\ \hline
		FAFA1 & 28 & 0\% & 41 & 35.94\% & 306.464 $\mu$J & 36.867\% \\ \hline
		FAFA2 & 28 & 0\% & 39 & 39.06\% & 287 $\mu$J & 40.877\% \\ \hline
		\multicolumn{7}{|c|}{$2^{nd}~scenario:~the~three~MSBs~are~exact,~and~the~five~LSBs~are~approximate.$} \\ \hline
		SIAFA1 \cite{ref22} & 19 & 32.14\% & 106 & -65.62\% & NC\textsuperscript{*} & NC\textsuperscript{*} \\ \hline
		SIAFA2 \cite{ref22} & 19 & 32.14\% & 116 & -81.25\% & NC\textsuperscript{*} & NC\textsuperscript{*} \\ \hline
		SIAFA3 \cite{ref22} & 19 & 32.14\% & 106 & -65.62\% & NC\textsuperscript{*} & NC\textsuperscript{*} \\ \hline
		SIAFA4 \cite{ref22} & 19 & 32.14\% & 106 & -65.62\% & NC\textsuperscript{*} & NC\textsuperscript{*} \\ \hline
		SAFAN \cite{ref1} & 19 & 32.14\% & 101 & -57.81\% & NC\textsuperscript{*} & NC\textsuperscript{*} \\ \hline
		SAID1 \cite{rref4} & 19 & 32.14\% & 76 & -18.75\% & NC\textsuperscript{*} & NC\textsuperscript{*} \\ \hline
		SAID2 \cite{rref4} & 24 & 14.29\% & 96 & -50\% & NC\textsuperscript{*} & NC\textsuperscript{*} \\ \hline
		SINC \cite{rref3} & 19 & 32.14\% & 81 & -26.56\% & NC\textsuperscript{*} & NC\textsuperscript{*} \\ \hline
		SINC+ \cite{rref3} & 19 & 32.14\% & 84 & -31.25\% & NC\textsuperscript{*} & NC\textsuperscript{*} \\ \hline
		FAFA1 & 28 & 0\% & 35 & 45.32\% & 261.722 $\mu$J & 46.084\% \\ \hline
		FAFA2 & 28 & 0\% & 35 & 45.32\% & 237.392 $\mu$J & 51.096\% \\ \hline
	\end{tabular}}
	\label{tab8}
	\begin{flushleft} \textsuperscript{*} \footnotesize Not Calculated. \end{flushleft}
\end{table}

As shown in Table \ref{tab8}, the memristors' count in both approximate scenarios and implementation approaches equals the memristors’ count in the exact 8-bit structure \cite{ref16}. In both scenarios, the approximate 8-bit RCA proposed by applying the FAFA2 obtained the best results in terms of the number of cycles and energy consumption compared to the exact 8-bit RCA \cite{ref16}, and the approximate 8-bit RCA contains the first approach's approximate full adder. This improved approximate 8-bit RCA reduces the number of cycles and energy consumption by up to 45.32\% and 51.096\%, respectively.

The number of memristors and computational steps of the proposed circuits in the structures of $1^{st}$ and $2^{nd}$ scenarios are also compared with the IMPLY-based exact and approximate circuits \cite{ref1,ref21,ref22,rref1,rref2,rref3,rref4} in Table \ref{tab8}. As mentioned earlier, it is impossible to directly compare the energy consumption of the proposed circuits with similar structures in \cite{ref1,ref21,ref22,rref1,rref2,rref3,rref4} due to the difference in the memristor model applied. The number of memristors required in the $1^{st}$ and $2^{nd}$ implementation scenarios structures, based on the exact and approximate circuits proposed in \cite{ref1,ref21,ref22,rref1,rref2,rref3,rref4}, is 14.29\%-32.14\% less than the FELIX-based exact and approximate structures. Suppose it is necessary to store the inputs in the work memristors using the IMPLY design method and reuse them later. In that case, the number of memristors required in the exact and approximate 8-bit structures ($1^{st}$ and $2^{nd}$ scenarios) will be equal to 36, which is 8 memristors more than the number of memristors required in the FELIX-based exact and approximate 8-bit adders. The number of computational steps of the proposed circuits in the aforementioned dual scenarios is at most 80.11\% and at least 46.05\% less than the exact and approximate 8-bit structures based on IMPLY-based approximate full adders \cite{ref1,ref21,ref22,rref1,rref2,rref3,rref4}. The improvement percentages reported in Table \ref{tab8} do not consider the computational steps of copying input data into the surplus memristors in the IMPLY method.

\subsection{Error-Analysis Results} \label{sec42}
In designing approximate arithmetic circuits, reducing hardware complexity and delay achieves this at the cost of losing an acceptable amount of accuracy. Hence, metrics such as MED and NMED need to be applied to evaluate the proposed approximate designs. For this reason, the FAFA has been evaluated in both scenarios using the error evaluation criteria. For this purpose, all the input states (65536 possible states) have been applied to the proposed circuit in both scenarios, and the error evaluation criteria, such as MED and NMED, have been calculated. It should be mentioned that, considering that the FAFA1 and FAFA2 result in the same values and truth table, the MED and NMED of both approaches are the same. The error evaluation metrics of the approximate RCAs (two scenarios) are written in Table \ref{tab9}.

In the proposed approximate full adder (FAFA), the $C_{out}$ is calculated precisely, which means that the error does not propagate from the least significant approximate full adders to more significant approximate and exact full adders. Accordingly, and based on the numbers reported in Table \ref{tab9}, the computational accuracy of the proposed circuit in both scenarios is higher than the computational accuracy of the 8-bit approximate arithmetic structures based on the IMPLY-based approximate full adders proposed in \cite{ref1,ref22,rref4} and the SINC design in \cite{rref3} in $1^{st}$ and $2^{nd}$ scenarios.

\begin{table}[!]
	\centering
	\caption{Error analysis results of the approximate 8-bit RCAs in two different scenarios.}
	\scalebox{1}{
	\begin{tabular}{|c|c|c|}
		\hline
		8-bit Serial Approximate RCA & MED & NMED \\ \hline
		\multicolumn{3}{|c|}{$1^{st}~scenario:~the~four~MSBs~are~exact,~and~the~four~LSBs~are~approximate.$} \\ \hline
		SIAFA1 \cite{ref22} & 4.351 & 0.0085  \\ \hline
		SIAFA2 \cite{ref22} & 6.1718 & 0.0121  \\ \hline
		SIAFA3 \cite{ref22} & 4.351 & 0.0085 \\ \hline
		SIAFA4 \cite{ref22} & 5.3125 & 0.0104  \\ \hline
		SAFAN \cite{ref1} & 5.78125 & 0.0113  \\ \hline
		SAID1 \cite{rref4} & 5.3125 & 0.0104  \\ \hline
		SAID2 \cite{rref4} & 4.3047 & 0.0084  \\ \hline
		SINC \cite{rref3} & 3.75 & 0.0074  \\ \hline
		SINC+ \cite{rref3} & 2.875 & 0.0056  \\ \hline
		FAFA1 and FAFA2 & 3.617 & 0.007 \\ \hline
		\multicolumn{3}{|c|}{$2^{nd}~scenario:~the~three~MSBs~are~exact,~and~the~five~LSBs~are~approximate.$} \\ \hline
		SIAFA1 \cite{ref22} & 8.8554 & 0.0173  \\ \hline
		SIAFA2 \cite{ref22} & 13.498 & 0.0264  \\ \hline
		SIAFA3 \cite{ref22} & 8.8554 & 0.0173 \\ \hline
		SIAFA4 \cite{ref22} & 10.6562 & 0.0208  \\ \hline
		SAFAN \cite{ref1} & 11.04687 & 0.02166  \\ \hline
		SAID1 \cite{rref4} & 10.6562 & 0.0209  \\ \hline
		SAID2 \cite{rref4} & 8.5293 & 0.0167  \\ \hline
		SINC \cite{rref3} & 7.75 & 0.0152  \\ \hline
		SINC+ \cite{rref3} & 5.875 & 0.0115  \\ \hline
		FAFA1 and FAFA2 & 7.376 & 0.014 \\ \hline
	\end{tabular}}
	\label{tab9}
\end{table}

\subsection{Simulations Of Various Image Processing Applications} \label{sec43}
Approximate computing can be applied in some applications that depend on human senses because human vision and hearing abilities are limited \cite{ref39}. Approximate adders can be employed in error-resilient applications such as image processing. In this paper, the FAFA has been evaluated in four different image processing applications. These four applications are image addition, grayscale filter, motion detection, and average pooling.

To evaluate the FAFA in the image addition application, all the corresponding pixels of two 256$\times$256 8-bit grayscale images are added together, and the result of the addition is stored in the corresponding pixel of the output image.

In the motion detection application, all the corresponding pixels of two 512$\times$512 pixel grayscale images \cite{ref40} are subtracted from each other, and the subtraction result is saved in the corresponding pixel of the output image, which shows the movement of two objects in two sequential frames \cite{ref1,ref22,ref38}.

To evaluate the FAFA by applying the grayscale filter, the R, G, and B values of each pixel were first added together. Then, the result of this addition was divided by 3 using an exact divider. The final result was written in the corresponding pixel of the output image \cite{ref1,ref22,ref38}. For this purpose, an 8-bit RGB image of 684$\times$912 pixels was chosen and converted into a same-sized grayscale photo. 

With the expansion of neural network use across various processing applications, Convolutional Neural Networks (CNNs) are among the most popular neural network types in image processing \cite{pooling}. One of the effective computational operations in CNNs is pooling, which reduces the input dimensions. Using the pooling function increases computational efficiency while preserving the main features of the input \cite{pooling}. One pooling technique used in CNNs is average pooling, which reduces the effect of input noise \cite{pooling}. To implement this operation, a 2$\times$2 kernel is sliding on the input image and the average value of every 4 pixels is stored as the final result. If the input of this process is an $n \times n$ image, the output is an $\frac{n}{2} \times \frac{n}{2}$ image with the features of the original image preserved. The proposed approximate full adders are used in the approximate 8-bit RCAs of the $1^{st}$ and $2^{nd}$ scenarios to compute the final sum of the average pooling function.

\begin{figure}[!]
     \centering
     \begin{subfigure}[]{0.3\textwidth}
         \centering
         \includegraphics[width=0.8\textwidth]{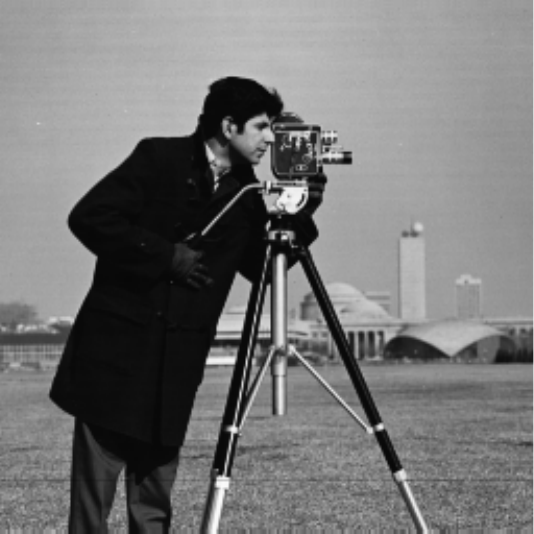}
         \caption{}
         \label{Fig8a}
     \end{subfigure}
     \hfill
     \begin{subfigure}[]{0.3\textwidth}
         \centering
         \includegraphics[width=0.8\textwidth]{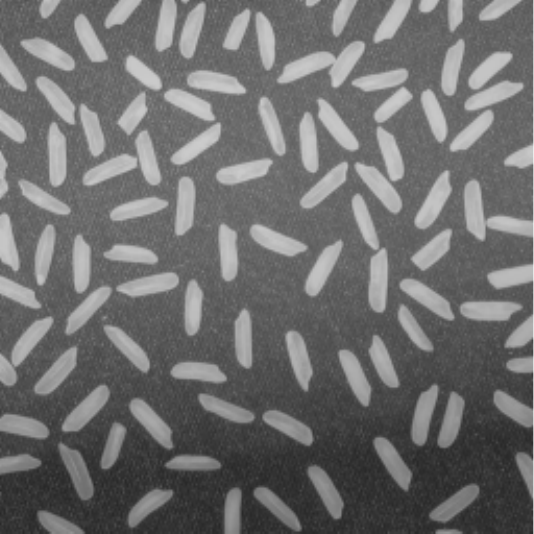}
         \caption{}
         \label{Fig8b}
     \end{subfigure}
     \hfill
     \begin{subfigure}[]{0.3\textwidth}
         \centering
         \includegraphics[width=0.8\textwidth]{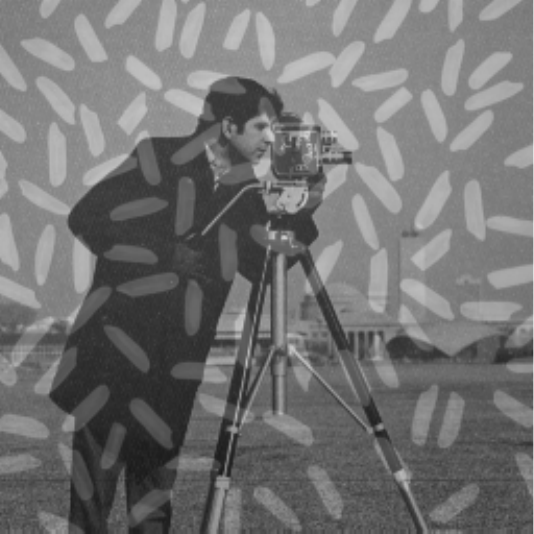}
         \caption{}
         \label{Fig8c}
     \end{subfigure}
     \begin{subfigure}[]{0.3\textwidth}
         \centering
         \includegraphics[width=0.8\textwidth]{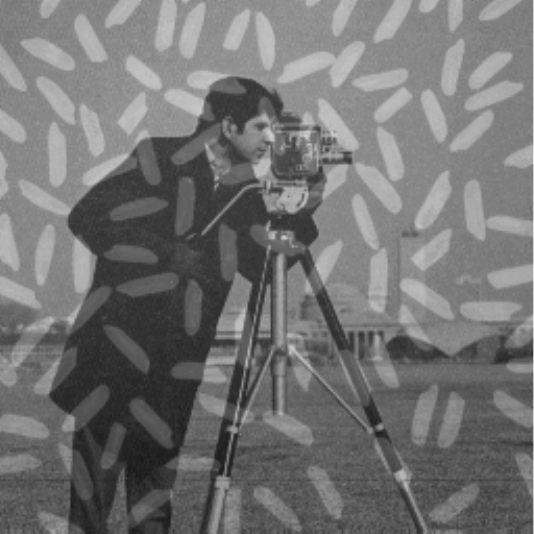}
         \caption{}
         \label{Fig8d}
     \end{subfigure}
     \begin{subfigure}[]{0.3\textwidth}
         \centering
         \includegraphics[width=0.8\textwidth]{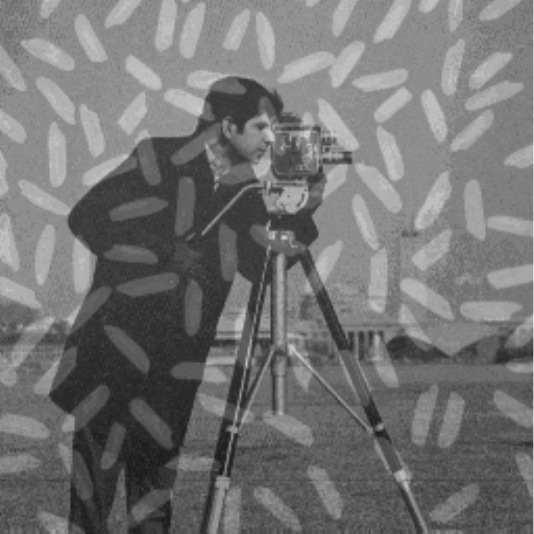}
         \caption{}
         \label{Fig8e}
     \end{subfigure}
        \caption{The results of image addition application: (a) cameraman, (b) rice, (c) exact addition, (d) approximate addition in the $1^{st}$ scenario, and (e) approximate addition in the $2^{nd}$ scenario (quantitative metrics are listed in Table \ref{tab10}).}
        \label{Fig8}
\end{figure}

\begin{figure}[tbp]
     \centering
     \begin{subfigure}[]{0.3\textwidth}
         \centering
         \includegraphics[width=0.8\textwidth]{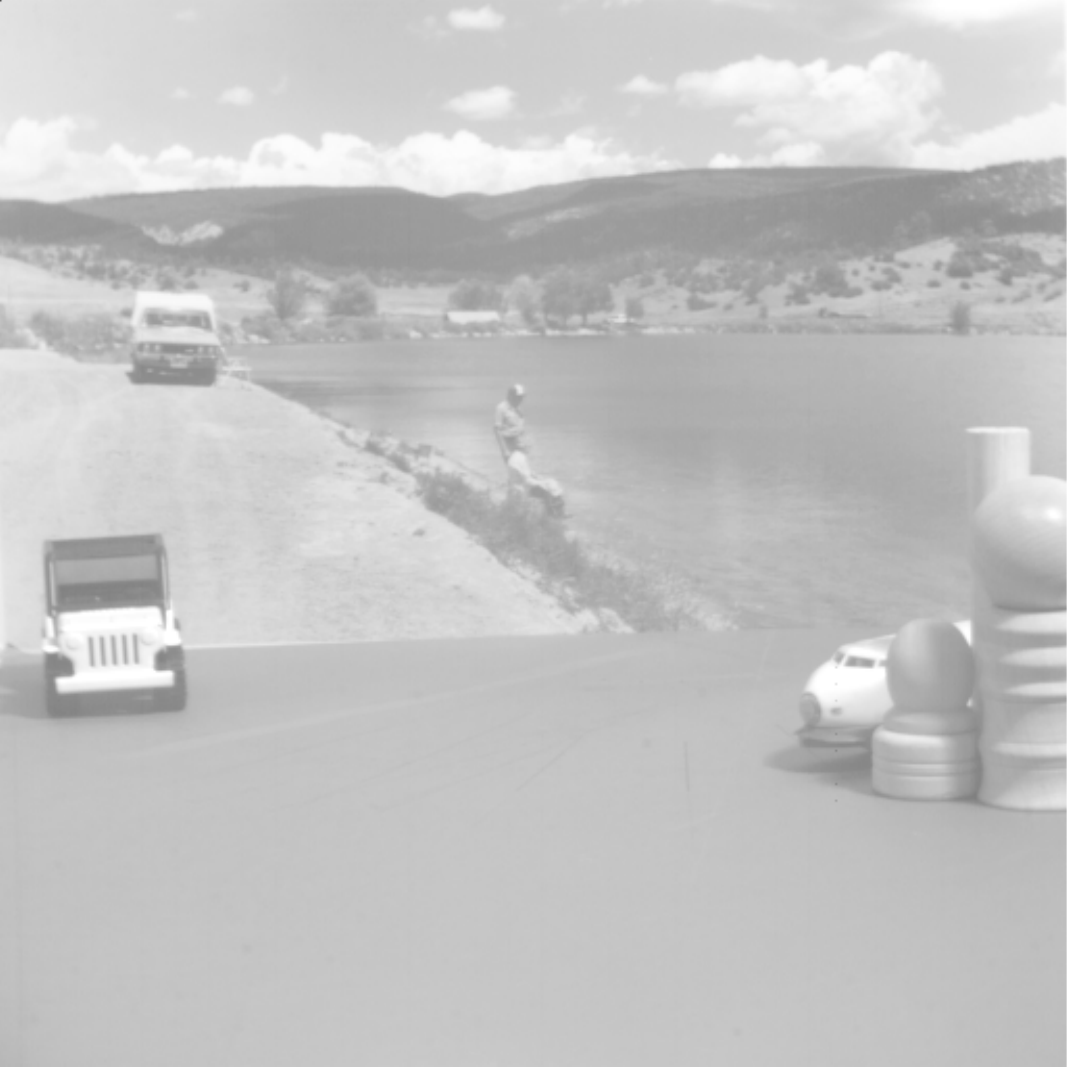}
         \caption{}
         \label{Fig9a}
     \end{subfigure}
     \hfill
     \begin{subfigure}[]{0.3\textwidth}
         \centering
         \includegraphics[width=0.8\textwidth]{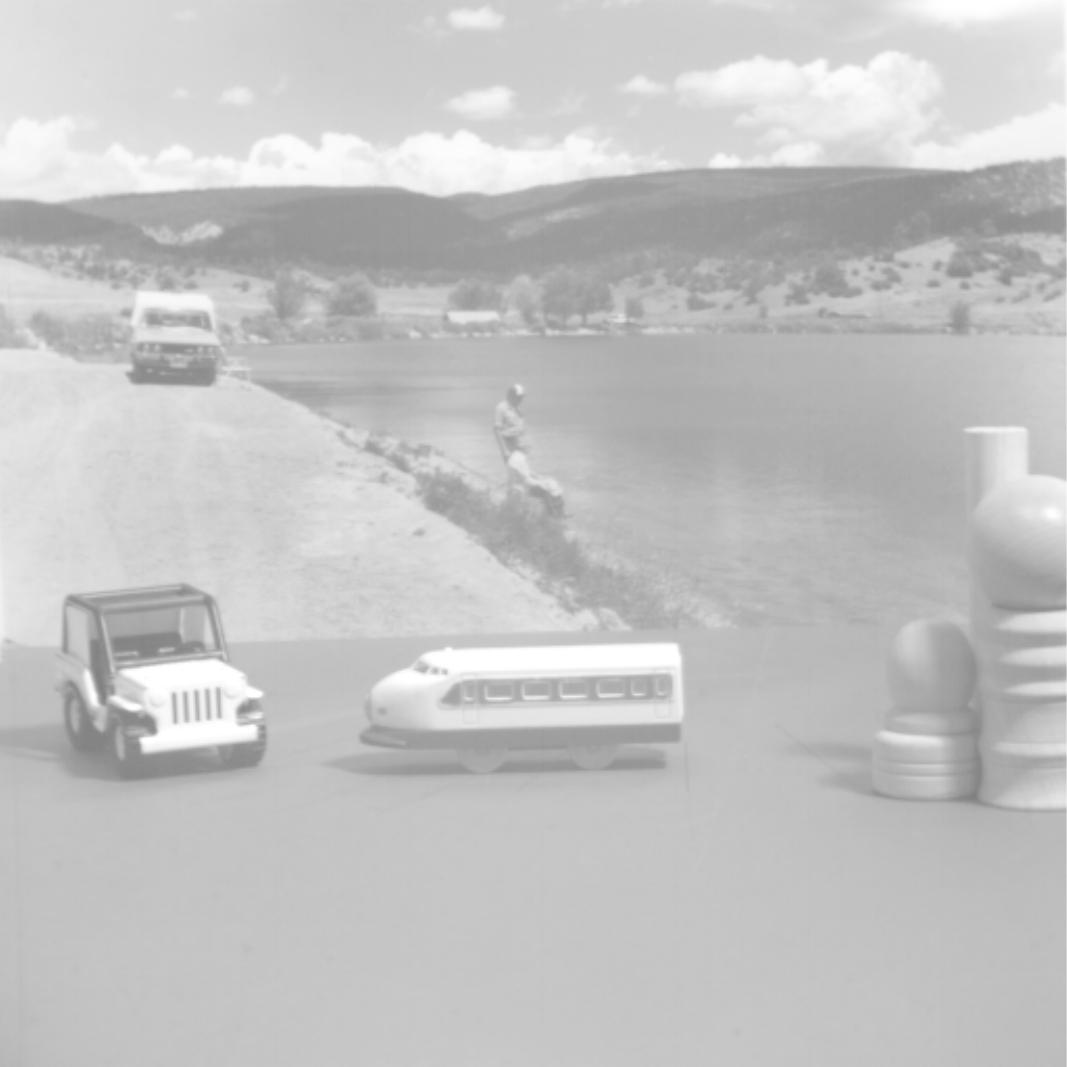}
         \caption{}
         \label{Fig9b}
     \end{subfigure}
     \hfill
     \begin{subfigure}[]{0.3\textwidth}
         \centering
         \includegraphics[width=0.8\textwidth]{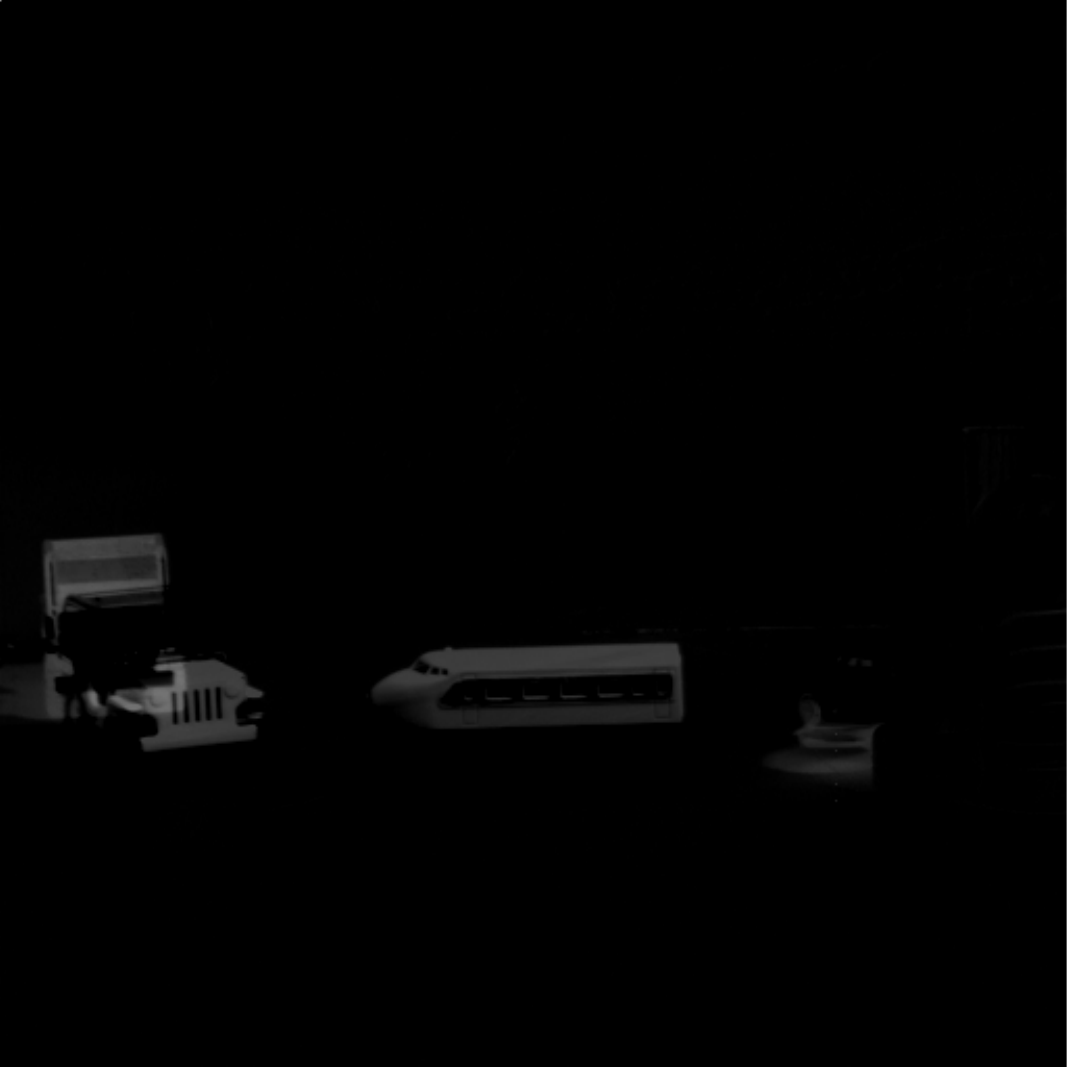}
         \caption{}
         \label{Fig9c}
     \end{subfigure}
     \begin{subfigure}[]{0.3\textwidth}
         \centering
         \includegraphics[width=0.8\textwidth]{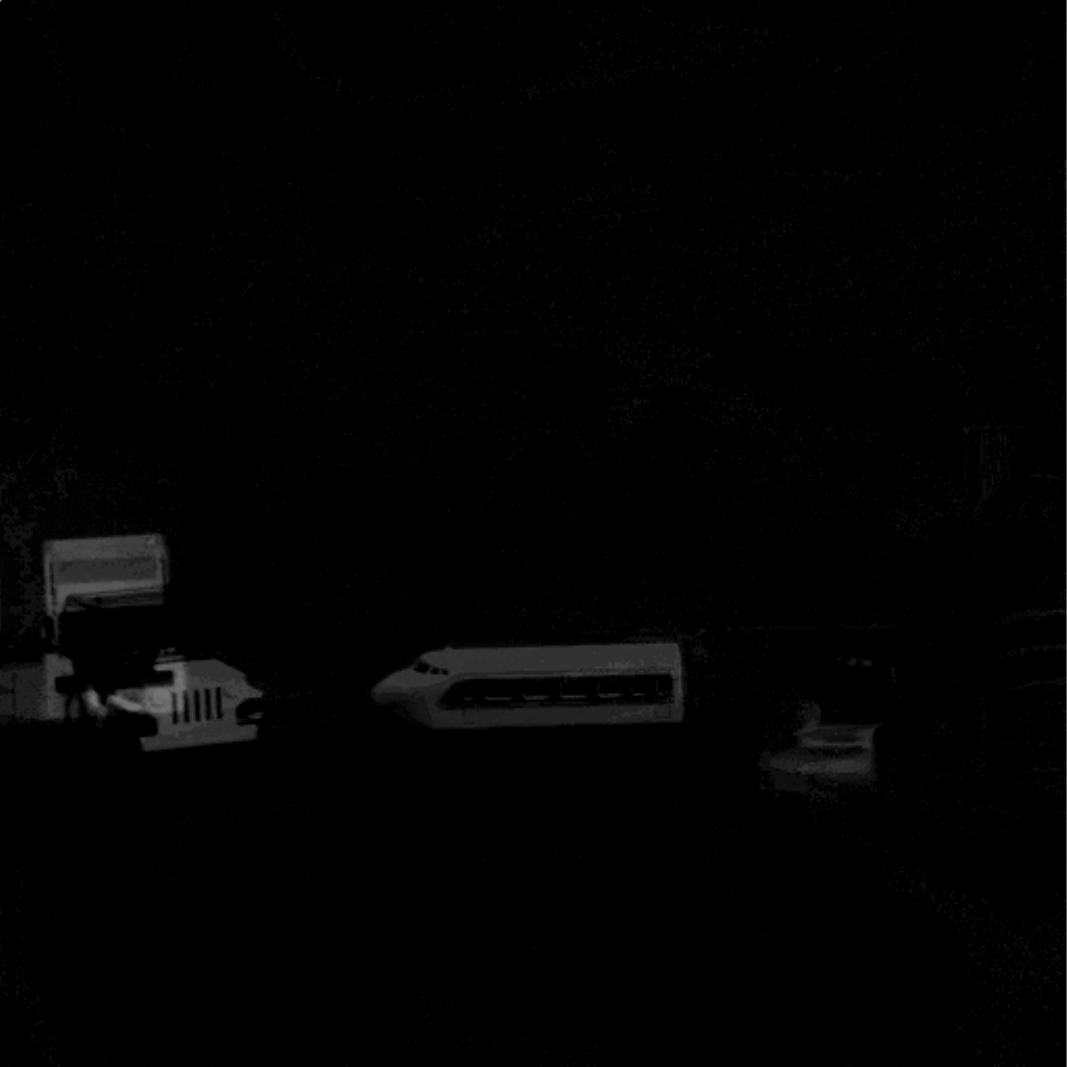}
         \caption{}
         \label{Fig9d}
     \end{subfigure}
     \begin{subfigure}[]{0.3\textwidth}
         \centering
         \includegraphics[width=0.8\textwidth]{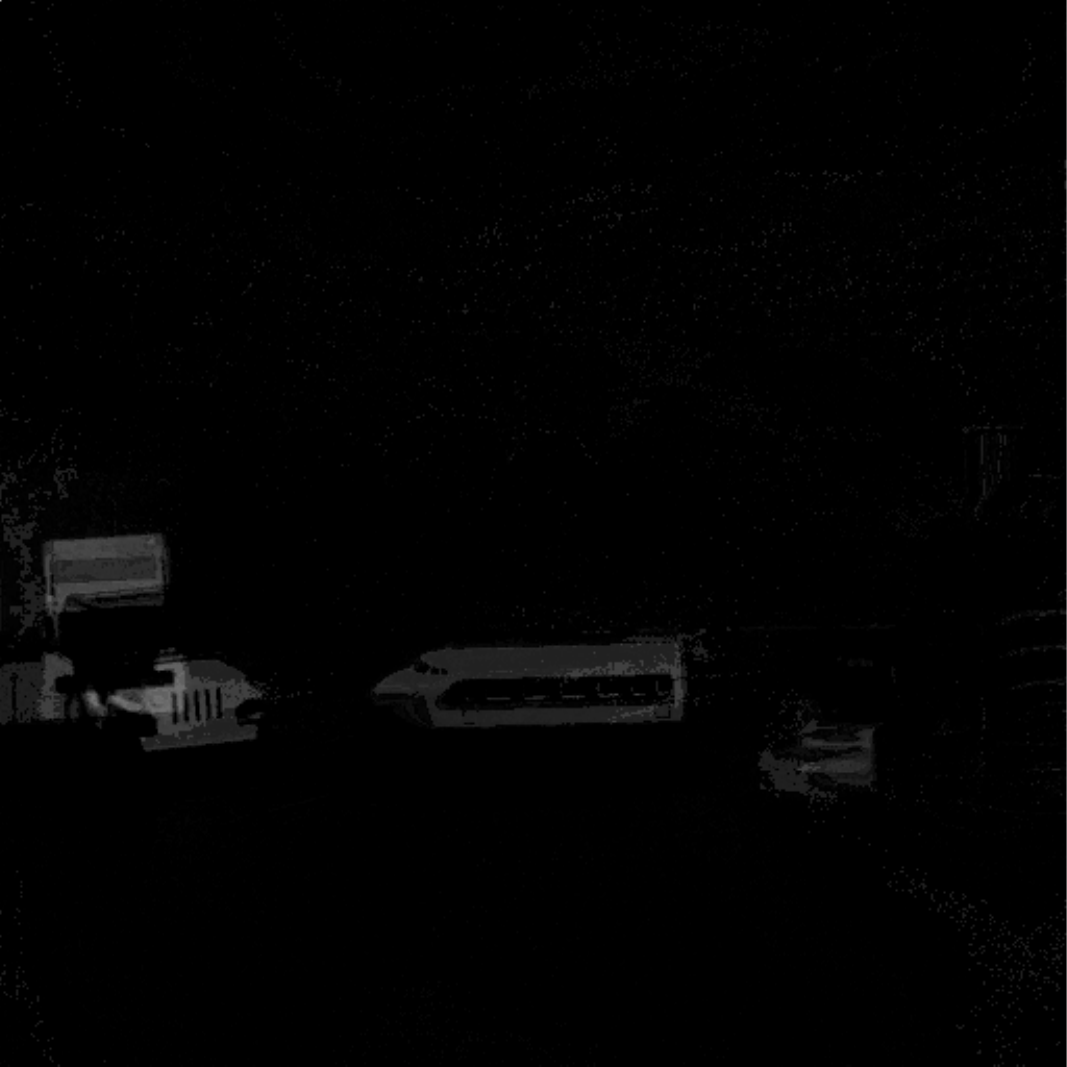}
         \caption{}
         \label{Fig9e}
     \end{subfigure}
        \caption{Motion detection application results: (a) first image, (b) second image, (c) exact result, (d) approximate motion detection in the $1^{st}$ scenario, and (e) approximate motion detection in the $2^{nd}$ scenario (quantitative metrics are listed in Table \ref{tab11}).}
        \label{Fig9}
\end{figure}

\begin{figure}[tbp]
     \centering
     \begin{subfigure}[]{0.45\textwidth}
         \centering
         \includegraphics[width=0.8\textwidth]{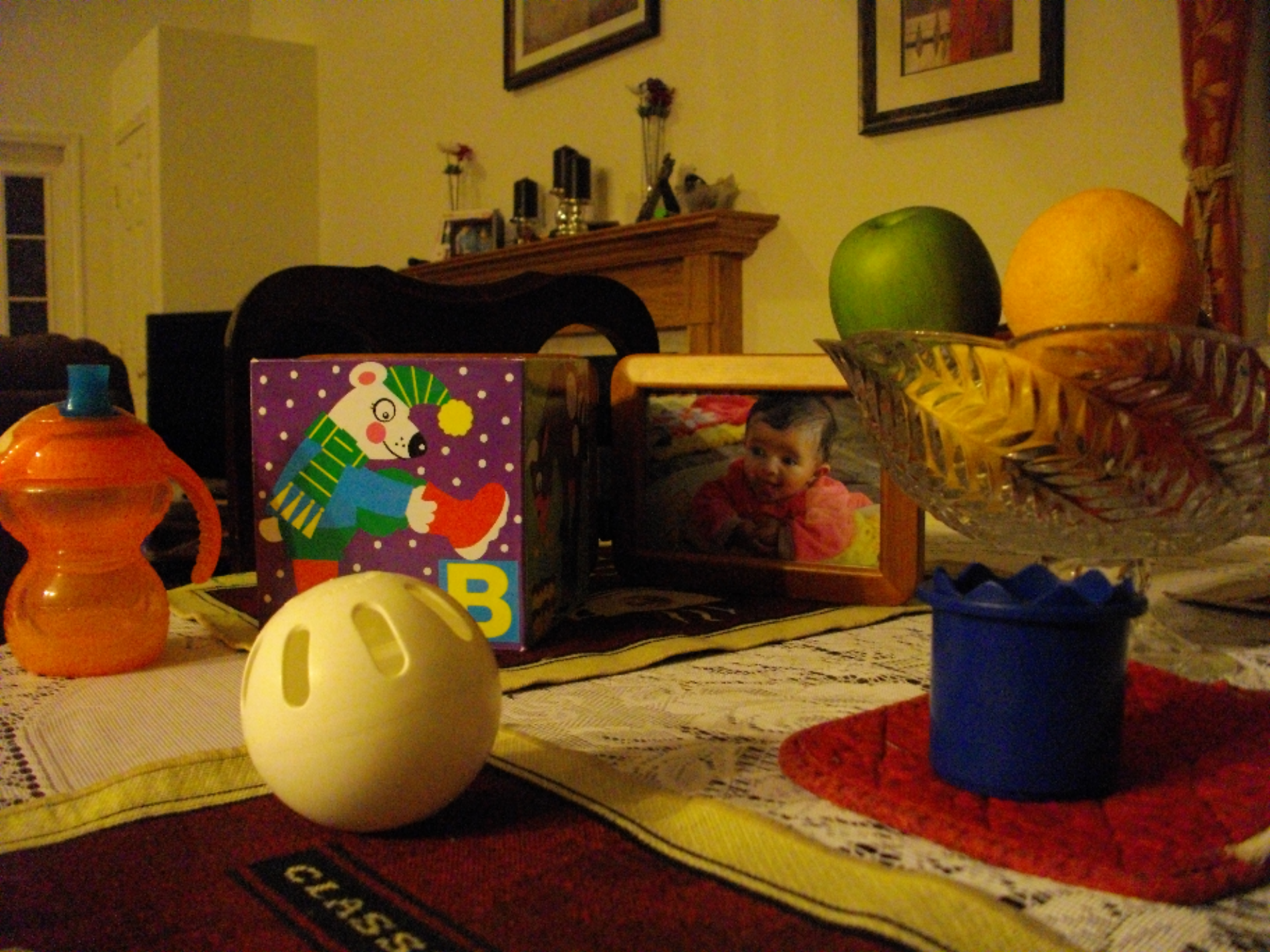}
         \caption{}
         \label{Fig10a}
     \end{subfigure}
     \begin{subfigure}[]{0.45\textwidth}
         \centering
         \includegraphics[width=0.8\textwidth]{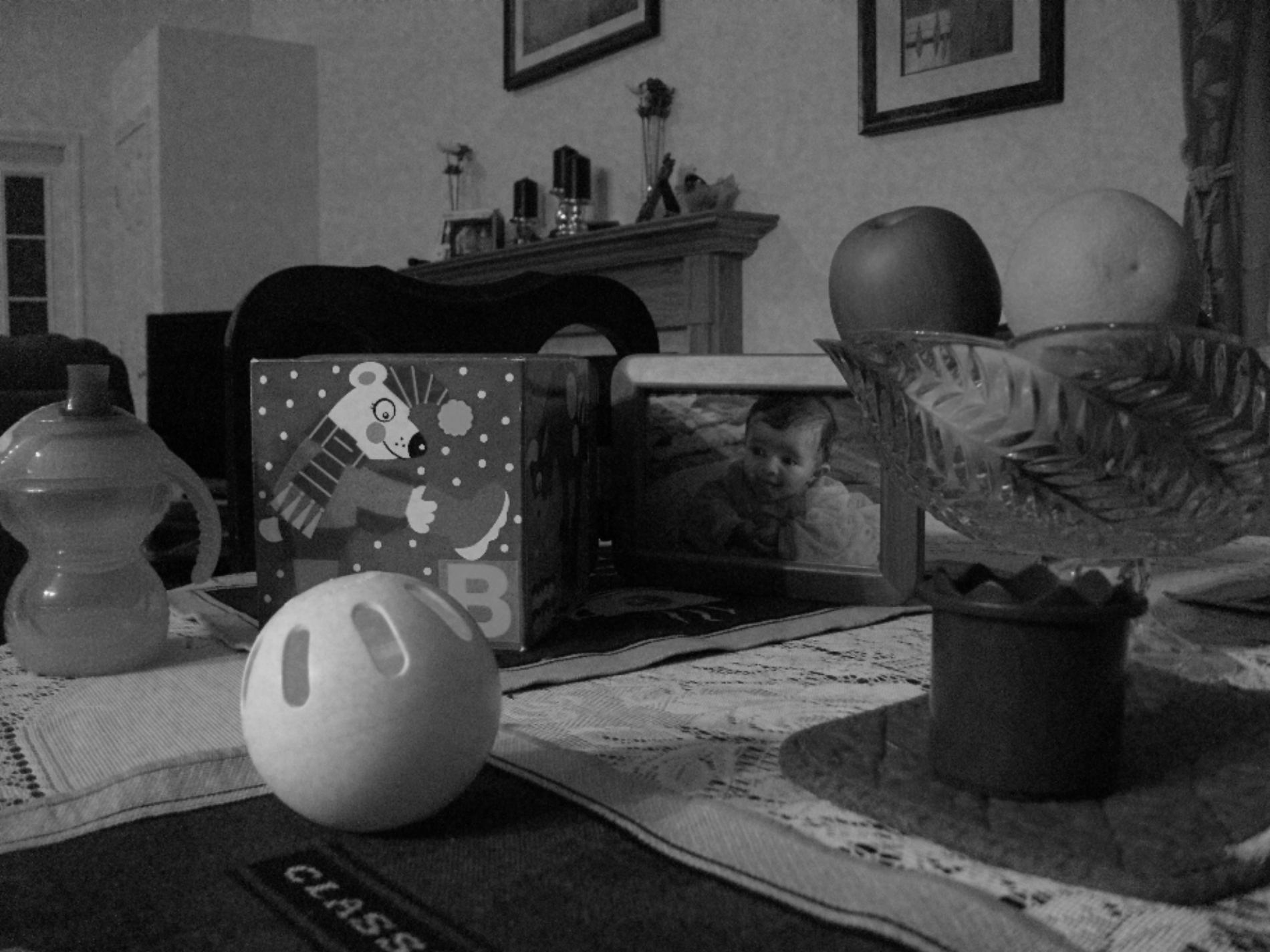}
         \caption{}
         \label{Fig10b}
     \end{subfigure}
     \begin{subfigure}[]{0.45\textwidth}
         \centering
         \includegraphics[width=0.8\textwidth]{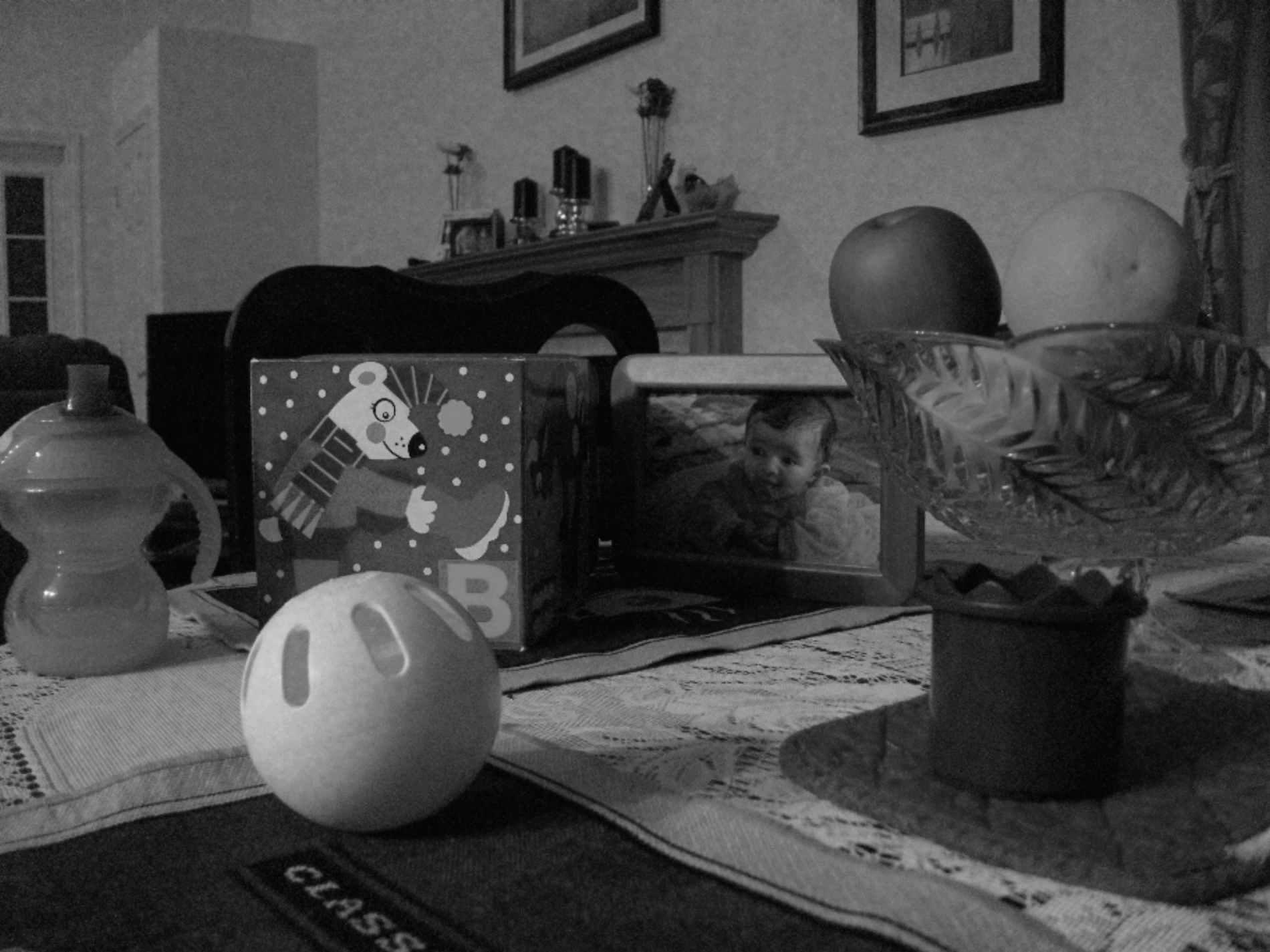}
         \caption{}
         \label{Fig10c}
     \end{subfigure}
     \begin{subfigure}[]{0.45\textwidth}
         \centering
         \includegraphics[width=0.8\textwidth]{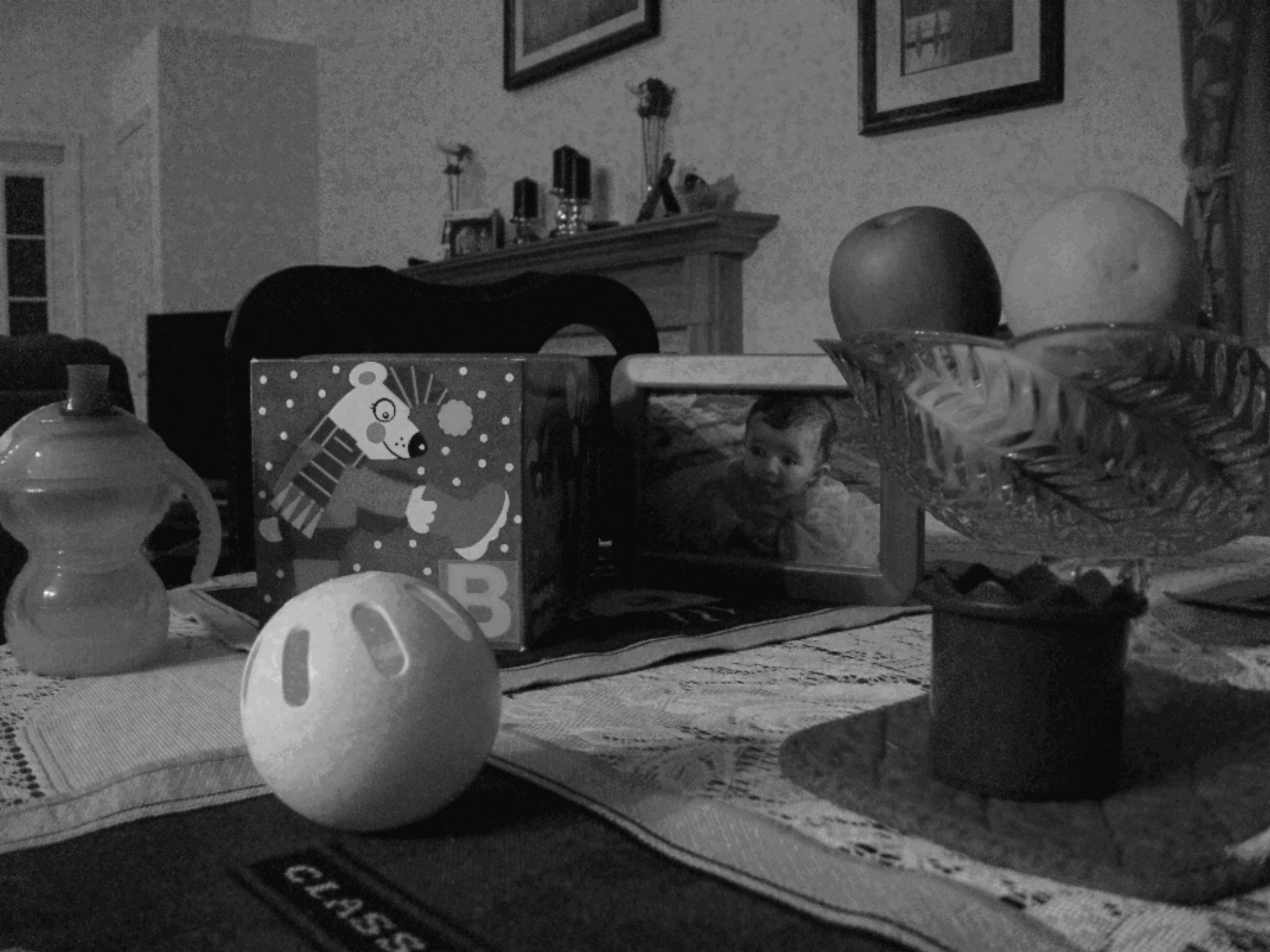}
         \caption{}
         \label{Fig10d}
     \end{subfigure}
        \caption{The results of applying the grayscale filter: (a) reference RGB image, (b) exact result, (c) approximate grayscale filter in the $1^{st}$ scenario, and (d) approximate grayscale filter in the $2^{nd}$ scenario (quantitative metrics are listed in Table \ref{tab12}).}
        \label{Fig10}
\end{figure}

\begin{figure}[tbp]
     \centering
     \begin{subfigure}[]{\textwidth}
         \centering
         \includegraphics[width=0.3\textwidth]{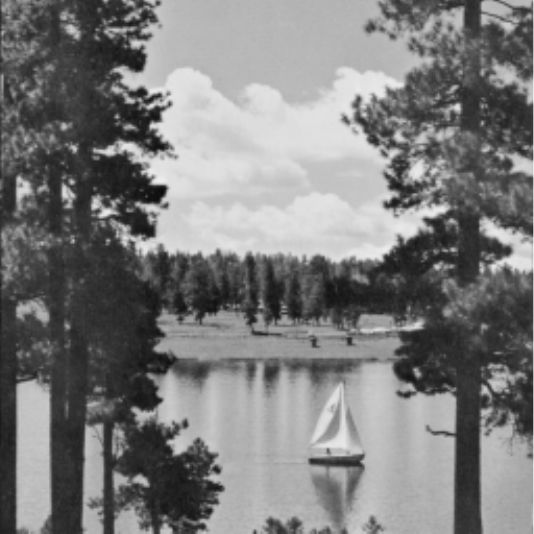}
         \caption{}
         \label{Fig11a}
     \end{subfigure}
     
     \begin{subfigure}[]{0.3\textwidth}
         \centering
         \includegraphics[width=0.5\textwidth]{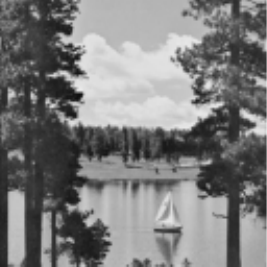}
         \caption{}
         \label{Fig11b}
     \end{subfigure}
     \hfill
     \begin{subfigure}[]{0.3\textwidth}
         \centering
         \includegraphics[width=0.5\textwidth]{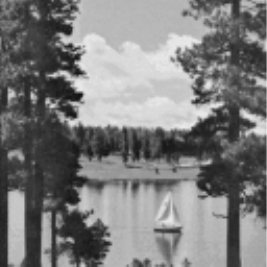}
         \caption{}
         \label{Fig11c}
     \end{subfigure}
     \hfill
     \begin{subfigure}[]{0.3\textwidth}
         \centering
         \includegraphics[width=0.5\textwidth]{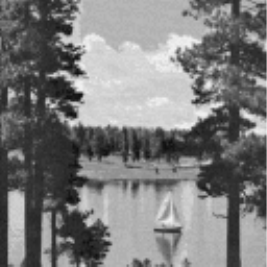}
         \caption{}
         \label{Fig11d}
     \end{subfigure}
        \caption{Average pooling application results: (a) reference grayscale image, (b) exact result, (c) approximate average pooling in the $1^{st}$ scenario, and (e) approximate average pooling in the $2^{nd}$ scenario (quantitative metrics are listed in Table \ref{tab13}).}
        \label{Fig11}
\end{figure}

The simulation results of these applications are written in Tables \ref{tab10}-\ref{tab13}. Peak Signal-to-Noise Ratio (PSNR), Structural Similarity Index Measure (SSIM), and Mean SSIM (MSSIM) criteria have been used to assess the quality of output images resulting from these applications. More details about calculating these criteria are explained in \cite{ref7,ref41}. It is mentioned in \cite{ref42} that when the PSNR of the output images is greater than 30 dB, the quality of the output images is acceptable. As seen in Tables \ref{tab10}-\ref{tab13}, the FAFA in both scenarios and all four applications has achieved PSNR greater than 30 dB, indicating the acceptable quality of the output images. Simulation outputs with SSIM values greater than 0.9 are acceptable for image processing applications \cite{ref43}. Our simulation results in both scenarios achieve SSIM values greater than 0.9 in all cases except the second scenario in the motion detection application. Also, MSSIM values in all applications and scenarios are greater than 0.9, which is acceptable for output images. The simulation results of these four applications applying FAFA in both scenarios are shown in Figures \ref{Fig8}-\ref{Fig11}.

\begin{table}[!]
	\centering
	\caption{Image quality evaluation metrics in the image addition application.}
	\scalebox{0.9}{
	\begin{tabular}{|c|c|c|c|}
		\hline
		8-bit Serial Approximate RCA & PSNR (dB) & SSIM & MSSIM \\ \hline
		\multicolumn{4}{|c|}{$1^{st}~scenario:~the~four~MSBs~are~exact,~and~the~four~LSBs~are~approximate.$} \\ \hline
		SIAFA1 \cite{ref22} & 38.67 & 0.9644 & 0.9649 \\ \hline
		SIAFA2 \cite{ref22} & 35.4576 & 0.9425 & 0.9436  \\ \hline
		SIAFA3 \cite{ref22} & 38.8399 &0.9638 & 0.9644 \\ \hline
		SIAFA4 \cite{ref22} & 37.8083 & 0.959 & 0.9597  \\ \hline
		SAFAN \cite{ref1} & 36.6395 & 0.9793 & 0.9796  \\ \hline
		SAID1 \cite{rref4} & 37.7197 & 0.9577 & 0.9582 \\ \hline
		SAID2 \cite{rref4} & 38.9996 & 0.9706 & 0.9712 \\ \hline
		SINC \cite{rref3} & 39.4496 & 0.9826 & 0.9826 \\ \hline
		SINC+ \cite{rref3} & 41.9405 & 0.9829 & 0.983 \\ \hline
		FAFA1 and FAFA2 & 39.471 & 0.97 & 0.97 \\ \hline
		\multicolumn{4}{|c|}{$2^{nd}~scenario:~the~three~MSBs~are~exact,~and~the~five~LSBs~are~approximate.$} \\ \hline
		SIAFA1 \cite{ref22} & 32.9823 & 0.8974 & 0.8996 \\ \hline
		SIAFA2 \cite{ref22} & 28.2504 & 0.8156 & 0.8166  \\ \hline
		SIAFA3 \cite{ref22} & 32.6497 &0.8905 & 0.8915 \\ \hline
		SIAFA4 \cite{ref22} & 32.0442 & 0.8931 & 0.8956  \\ \hline
		SAFAN \cite{ref1} & 30.5866 & 0.9297 & 0.9298  \\ \hline
		SAID1 \cite{rref4} & 32.062 & 0.8797 & 0.8819 \\ \hline
		SAID2 \cite{rref4} & 32.988 & 0.9077 & 0.9086 \\ \hline
		SINC \cite{rref3} & 34.0182 & 0.9519 & 0.9534 \\ \hline
		SINC+ \cite{rref3} & 36.2917 & 0.9522 & 0.9532 \\ \hline
		FAFA1 and FAFA2 & 33.776 & 0.912 & 0.914 \\ \hline
	\end{tabular}}
	\label{tab10}
\end{table}

\begin{table}[!]
	\centering
	\caption{Image quality evaluation criteria in the motion detection application.}
	\scalebox{0.9}{
	\begin{tabular}{|c|c|c|c|}
		\hline
		8-bit Serial Approximate RCA & PSNR (dB) & SSIM & MSSIM \\ \hline
		\multicolumn{4}{|c|}{$1^{st}~scenario:~the~four~MSBs~are~exact,~and~the~four~LSBs~are~approximate.$} \\ \hline
		SIAFA1 \cite{ref22} & 37.4131 & 0.6405 & 0.6705 \\ \hline
		SIAFA2 \cite{ref22} & 37.5605 & 0.9338 & 0.9652  \\ \hline
		SIAFA3 \cite{ref22} & 36.8613 & 0.5993 & 0.6302 \\ \hline
		SIAFA4 \cite{ref22} & 40.3861 & 0.9131 & 0.9423  \\ \hline
		SAFAN \cite{ref1} & 44.0205 & 0.9798 & 0.9857  \\ \hline
		SAID1 \cite{rref4} & 36.8708 & 0.7056 & 0.7299 \\ \hline
		SAID2 \cite{rref4} & 36.6587 & 0.6387 & 0.6505 \\ \hline
		SINC \cite{rref3} & 38.8525 & 0.7159 & 0.7393 \\ \hline
		SINC+ \cite{rref3} & 42.9661 & 0.8071 & 0.8177 \\ \hline
		FAFA1 and FAFA2 & 40.788 & 0.93 & 0.958 \\ \hline
		\multicolumn{4}{|c|}{$2^{nd}~scenario:~the~three~MSBs~are~exact,~and~the~five~LSBs~are~approximate.$} \\ \hline
		SIAFA1 \cite{ref22} & 32.6121 & 0.508 & 0.5404 \\ \hline
		SIAFA2 \cite{ref22} & 31.6441 & 0.8991 & 0.9265  \\ \hline
		SIAFA3 \cite{ref22} & 32.4096 & 0.4747 & 0.5094 \\ \hline
		SIAFA4 \cite{ref22} & 35.0436 & 0.8664 & 0.902  \\ \hline
		SAFAN \cite{ref1} & 37.5336 & 0.9667 & 0.9727  \\ \hline
		SAID1 \cite{rref4} & 30.914 & 0.5491 & 0.5508 \\ \hline
		SAID2 \cite{rref4} & 32.288 & 0.5679 & 0.5691 \\ \hline
		SINC \cite{rref3} & 35.1 & 0.6473 & 0.674 \\ \hline
		SINC+ \cite{rref3} & 39.8867 & 0.7445 & 0.7629 \\ \hline
		FAFA1 and FAFA2 & 35.309 & 0.887 & 0.922 \\ \hline
	\end{tabular}}
	\label{tab11}
\end{table}

\begin{table}[!]
	\centering
	\caption{Image quality evaluation criteria in grayscale filter application.}
	\scalebox{0.9}{
	\begin{tabular}{|c|c|c|c|}
		\hline
		8-bit Serial Approximate RCA & PSNR (dB) & SSIM & MSSIM \\ \hline
		\multicolumn{4}{|c|}{$1^{st}~scenario:~the~four~MSBs~are~exact,~and~the~four~LSBs~are~approximate.$} \\ \hline
		SIAFA1 \cite{ref22} & 41.4201 & 0.9693 & 0.9957 \\ \hline
		SIAFA2 \cite{ref22} & 35.9998 & 0.9263 & 0.9874  \\ \hline
		SIAFA3 \cite{ref22} & 41.2315 & 0.9684 & 0.9956 \\ \hline
		SIAFA4 \cite{ref22} & 36.9634 & 0.9451 & 0.989  \\ \hline
		SAFAN \cite{ref1} & 36.0101 & 0.9625 & 0.9886 \\ \hline
		SAID1 \cite{rref4} & 40.9647 & 0.9676 & 0.9956 \\ \hline
		SAID2 \cite{rref4} & 39.832 & 0.9629 & 0.9949 \\ \hline
		SINC \cite{rref3} & 36.6184 & 0.9564 & 0.9859 \\ \hline
		SINC+ \cite{rref3} & 41.1432 & 0.9655 & 0.9952 \\ \hline
		FAFA1 and FAFA2 & 41.906 & 0.973 & 0.996 \\ \hline
		\multicolumn{4}{|c|}{$2^{nd}~scenario:~the~three~MSBs~are~exact,~and~the~five~LSBs~are~approximate.$} \\ \hline
		SIAFA1 \cite{ref22} & 35.5671 & 0.9019 & 0.9778 \\ \hline
		SIAFA2 \cite{ref22} & 28.4883 & 0.7576 & 0.9317  \\ \hline
		SIAFA3 \cite{ref22} & 35.3588 & 0.8916 & 0.9794 \\ \hline
		SIAFA4 \cite{ref22} & 31.5146 & 0.8525 & 0.9589  \\ \hline
		SAFAN \cite{ref1} & 28.9472 & 0.8633 & 0.9482 \\ \hline
		SAID1 \cite{rref4} & 35.0164 & 0.9067 & 0.9723 \\ \hline
		SAID2 \cite{rref4} & 34.2532 & 0.8874 & 0.975 \\ \hline
		SINC \cite{rref3} & 31.1208 & 0.8962 & 0.9641 \\ \hline
		SINC+ \cite{rref3} & 35.2854 & 0.9047 & 0.9748 \\ \hline
		FAFA1 and FAFA2 & 35.864 & 0.909 & 0.981 \\ \hline
	\end{tabular}}
	\label{tab12}
\end{table}

\begin{table}[!]
	\centering
	\caption{Image quality evaluation criteria in average pooling application.}
	\scalebox{0.9}{
	\begin{tabular}{|c|c|c|c|}
		\hline
		8-bit Serial Approximate RCA & PSNR (dB) & SSIM & MSSIM \\ \hline
		\multicolumn{4}{|c|}{$1^{st}~scenario:~the~four~MSBs~are~exact,~and~the~four~LSBs~are~approximate.$} \\ \hline
		SIAFA1 \cite{ref22} & 34.7106 & 0.983 & 0.9824 \\ \hline
		SIAFA2 \cite{ref22} & 35.8146 & 0.9684 & 0.9673  \\ \hline
		SIAFA3 \cite{ref22} & 41.6697 & 0.9886 & 0.9882 \\ \hline
		SIAFA4 \cite{ref22} & 35.0342 & 0.9787 & 0.9772  \\ \hline
		SAFAN \cite{ref1} & 35.1589 & 0.9852 & 0.985 \\ \hline
		SAID1 \cite{rref4} & 39.6548 & 0.9802 & 0.979 \\ \hline
		SAID2 \cite{rref4} & 40.5835 & 0.986 & 0.9858 \\ \hline
		SINC \cite{rref3} & 34.7844 & 0.9835 & 0.9824 \\ \hline
		SINC+ \cite{rref3} & 40.5606 & 0.984 & 0.9833 \\ \hline

		FAFA1 and FAFA2 & 40.9997 & 0.9844 & 0.9838 \\ \hline
		\multicolumn{4}{|c|}{$2^{nd}~scenario:~the~three~MSBs~are~exact,~and~the~five~LSBs~are~approximate.$} \\ \hline
		SIAFA1 \cite{ref22} & 33.5434 & 0.9469 & 0.9435 \\ \hline
		SIAFA2 \cite{ref22} & 29.2172 & 0.8884 & 0.8825  \\ \hline
		SIAFA3 \cite{ref22} & 34.8578 & 0.9602 & 0.9575 \\ \hline
		SIAFA4 \cite{ref22} & 27.5895 & 0.9302 & 0.9236  \\ \hline
		SAFAN \cite{ref1} & 28.9039 & 0.9526 & 0.95 \\ \hline
		SAID1 \cite{rref4} & 32.9688 & 0.9397 & 0.9346 \\ \hline
		SAID2 \cite{rref4} & 34.7106 & 0.9625 & 0.9612 \\ \hline
		SINC \cite{rref3} & 27.2967 & 0.9453 & 0.9401 \\ \hline
		SINC+ \cite{rref3} & 33.4426 & 0.955 & 0.9522 \\ \hline
		FAFA1 and FAFA2 & 34.2609 & 0.9404 & 0.9374 \\ \hline
	\end{tabular}}
	\label{tab13}
\end{table}

\begin{figure}[!]
     \centering
     \begin{subfigure}[]{\textwidth}
         \centering
         \includegraphics[width=0.8\textwidth]{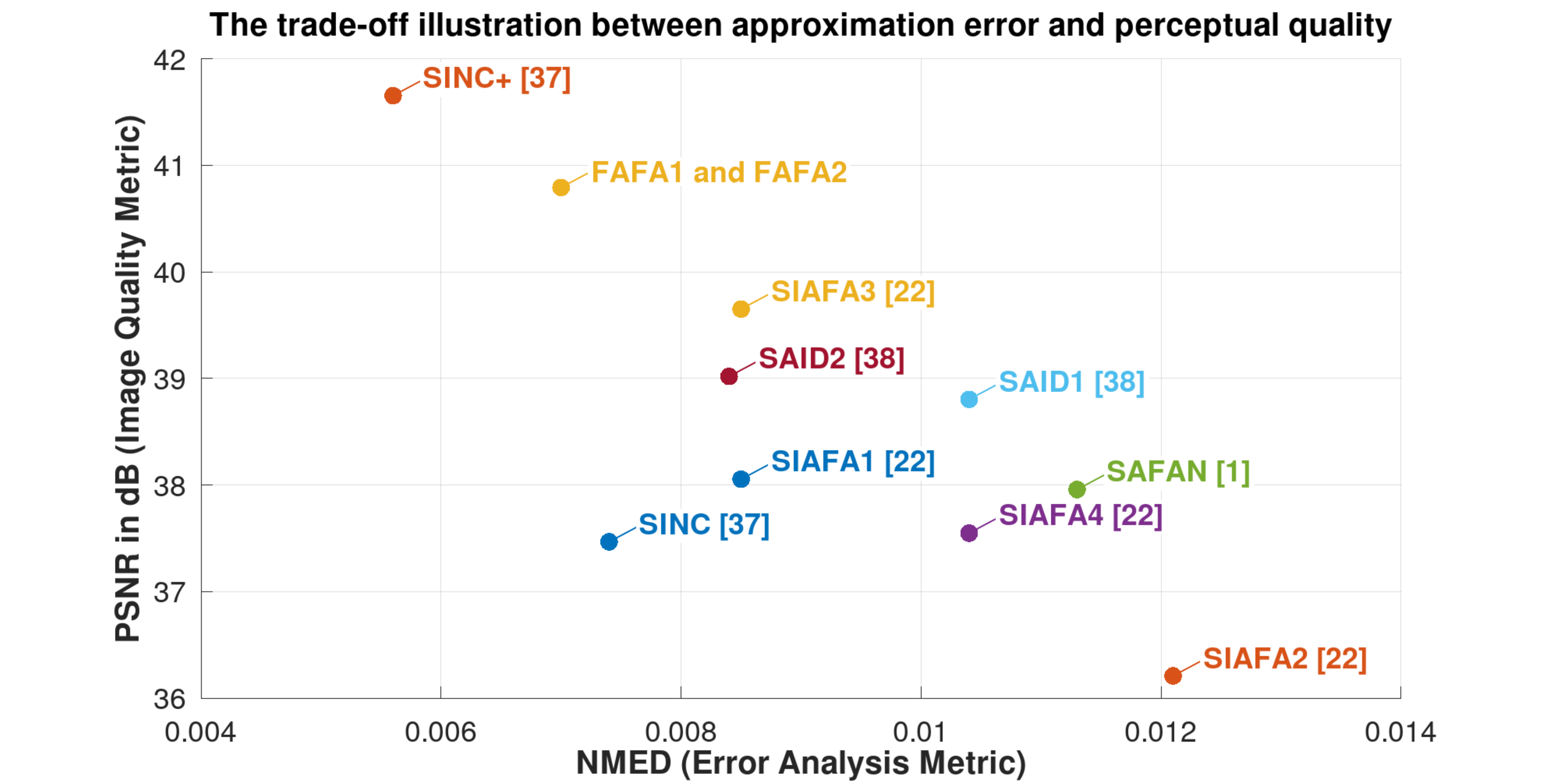}
         \caption{}
         \label{Fig12a}
     \end{subfigure}
     \begin{subfigure}[]{\textwidth}
         \centering
         \includegraphics[width=0.8\textwidth]{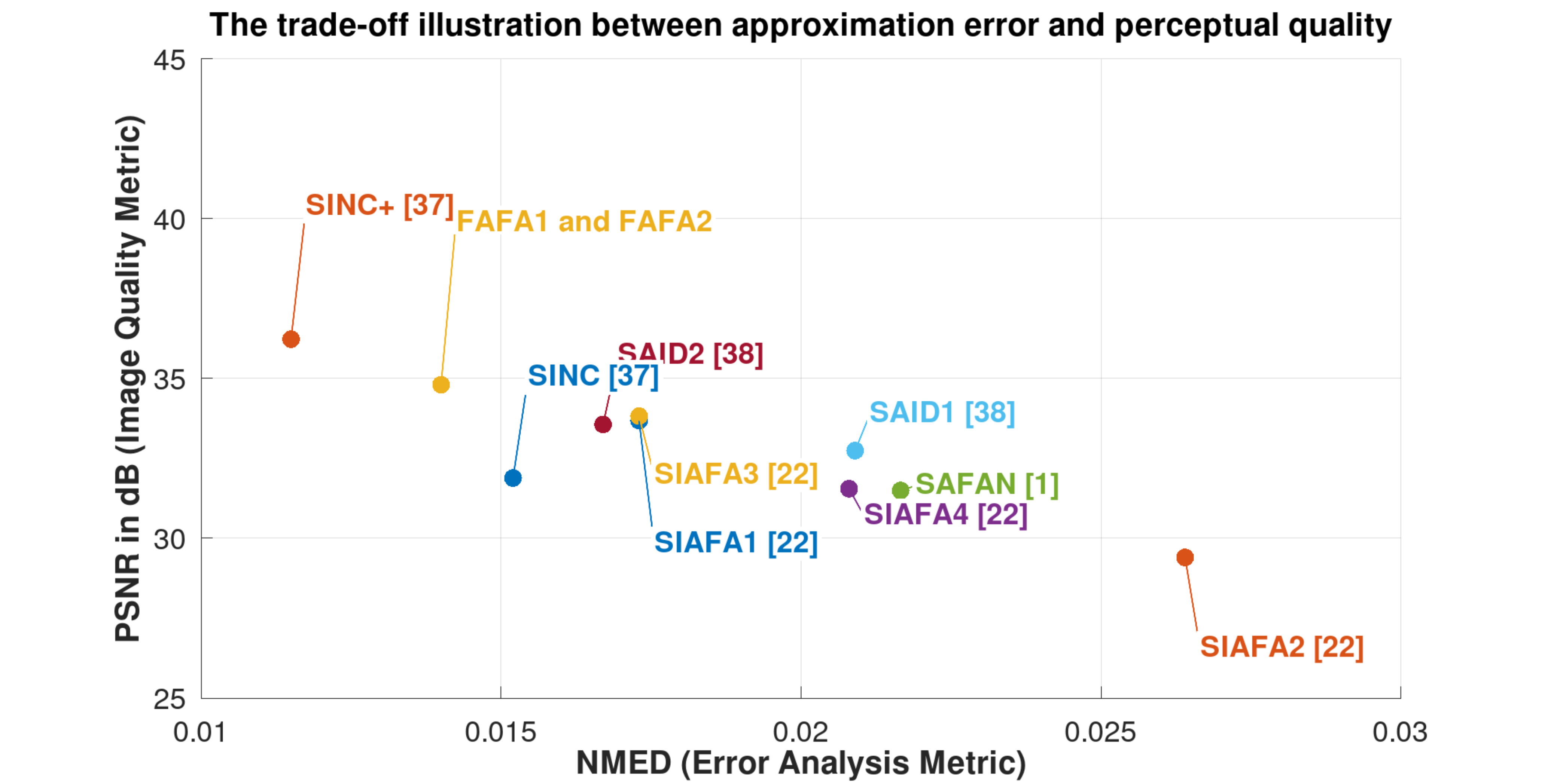}
         \caption{}
         \label{Fig12b}
     \end{subfigure}
        \caption{Image quality degradation (PSNR) versus error analysis metrics (NMED), the trade-off illustration between approximation error and perceptual quality: (a) $1^{st}$ scenario, and (b) $2^{nd}$ scenario.}
        \label{Fig12}
\end{figure}

According to the numbers reported in Tables \ref{tab10}-\ref{tab13}, the quality of output images in grayscale filter applications resulting from the arithmetic structures constructed of the proposed approximate full adder is superior to the other circuits \cite{ref1,ref22,rref3,rref4} in both scenarios, indicating that the computational accuracy has decreased to an acceptable level. In the image addition, motion detection, and average pooling applications, FAFA ranked among the top three in both scenarios, and its image quality met the criteria for visually acceptable images. The high quality of the output images indicates that the computational accuracy has decreased to an acceptable level, and the proposed circuit can be effectively employed in error-resilient computational applications.

In Figure \ref{Fig12}, two diagrams are drawn to examine the connection between the error metrics (NMED) and the image quality metrics of FAFA and state-of-the art \cite{ref1,ref22,rref3,rref4}, which are applied in the architectures of scenarios 1 and 2. The average PSNR of four applications was applied to draw this diagram. The results show that there is an inverse trend between these two evaluation criteria; the higher the NMED, the lower the image quality metric. By examining the diagrams in Figure \ref{Fig12}, it can be concluded that the proposed full adder cell and SINC+ \cite{rref3} are the best options for use in the error-resilient applications.

Considering the diagrams in Figure \ref{Fig12} and their purposes, it can be concluded that the approximate RCAs (in scenarios 1 and 2) implemented using SINC+ \cite{rref3} and FAFA have the highest computational accuracy. It is worth noting that the main goal of designing the FAFA was to improve the circuit evaluation criteria while maintaining acceptable computational accuracy, which has been achieved, as reported.

\section{Conclusion} \label{sec5}
This paper introduces a serial FELIX-based approximate full adder cell for IMC of error-resilient applications in two different implementation approaches. According to the advantages of approximate computing, the reduction of the number of memristors of the FAFA1 and FAFA2 compared to the exact FELIX-based full adder is 14.28\% and 28.57\%, respectively, and the energy consumption reduction is 73.735\% and 81.754\%, respectively. Also, the number of cycles in both approaches has decreased by 66.66\% compared to the exact design. In this paper, MED, NMED, PSNR, SSIM, and MSSIM criteria were used to evaluate the accuracy of the FAFA in two different scenarios and three different image processing and average pooling applications. In the $1^{st}$ and $2^{nd}$ scenarios, the exact full adder is applied in four and three MSBs; the FAFA adder is applied in four and five LSBs. The evaluations and simulations of the FAFA show that, in all applications and scenarios, PSNR is greater than 30 dB, and acceptable results were obtained regarding the accuracy criteria. In the mentioned scenarios, by applying the FAFA2, the number of cycles and energy consumption are improved by a maximum of 45.32\% and 51.096\%, respectively.

\appendix
\renewcommand{\thetable}{A\arabic{table}}
\setcounter{table}{0}

\section*{Appendix A: Application-level simulation-supplementary simulations and analyzes} \label{appendixa}
In subsection \ref{sec43}, the functionality of the proposed full adder cells (FAFA1 and FAFA2) was analyzed across various image processing applications, including average pooling, which is applicable in machine learning and CNNs. The applicability of the previous works \cite{ref1,ref22,rref3,rref4} was also examined in these applications. The generated outputs were evaluated using various criteria, and the results indicated that the proposed circuits can be efficiently applied to computational structures, such as RCAs, in scenarios 1 and 2.

To gain greater confidence in the performance of the proposed circuits in the aforementioned applications, the circuits were incorporated into 8-bit approximate adder structures in both scenarios, and their performance in each application was evaluated on a set of standard images from two datasets \cite{ref40,datasetfig}. The images in the dataset \cite{datasetfig} were used for image addition, grayscale filter, and average pooling, and three categories of images in the dataset \cite{ref40} were used for the motion detection application. Image quality evaluation criteria were calculated for each of the images in all four applications, and the average results obtained are reported separately in Tables \ref{taba1}-\ref{taba4} for each of the scenarios. By comparing the mentioned average results of the image quality metrics with those in Tables \ref{tab10}-\ref{tab13}, it can be concluded that the proposed circuits, when applied to the 8-bit approximate RCAs in scenarios 1 and 2, provide acceptable functionality across all four applications and produce outputs of acceptable quality. Therefore, considering the results reported in Tables \ref{tab10}-\ref{tab13} and \ref{taba1}-\ref{taba4}, it can be concluded that the use of the proposed approximate circuits, if used deliberately in larger arithmetic structures, improves the circuit evaluation criteria, such as energy consumption, and produces outputs with acceptable accuracy and appropriate quality. The ultimate goal of using approximate computing and designing approximate arithmetic circuits is also in accordance with this procedure and the results obtained.

\begin{table}[!]
	\centering
	\caption{Average Image quality metrics of the image addition application supplementary simulation based on the pictures of \cite{datasetfig}.}
	\scalebox{0.9}{
	\begin{tabular}{|c|c|c|c|}
		\hline
		8-bit Serial Approximate RCA & PSNR (dB) & SSIM & MSSIM \\ \hline
		\multicolumn{4}{|c|}{$1^{st}~scenario:~the~four~MSBs~are~exact,~and~the~four~LSBs~are~approximate.$} \\ \hline
		FAFA1 and FAFA2 & 39.445 & 0.969 & 0.97 \\ \hline
		\multicolumn{4}{|c|}{$2^{nd}~scenario:~the~three~MSBs~are~exact,~and~the~five~LSBs~are~approximate.$} \\ \hline
		FAFA1 and FAFA2 & 33.695 & 0.912 & 0.908 \\ \hline
	\end{tabular}}
	\label{taba1}
\end{table}

\begin{table}[!]
	\centering
	\caption{Average Image quality metrics of the motion detection application supplementary simulation based on the pictures of \cite{ref40}.}
	\scalebox{0.9}{
	\begin{tabular}{|c|c|c|c|}
		\hline
		8-bit Serial Approximate RCA & PSNR (dB) & SSIM & MSSIM \\ \hline
		\multicolumn{4}{|c|}{$1^{st}~scenario:~the~four~MSBs~are~exact,~and~the~four~LSBs~are~approximate.$} \\ \hline
		FAFA1 and FAFA2 & 39.712 & 0.948 & 0.957 \\ \hline
		\multicolumn{4}{|c|}{$2^{nd}~scenario:~the~three~MSBs~are~exact,~and~the~five~LSBs~are~approximate.$} \\ \hline
		FAFA1 and FAFA2 & 34.177 & 0.891 & 0.903 \\ \hline
	\end{tabular}}
	\label{taba2}
\end{table}

\begin{table}[!]
	\centering
	\caption{Average Image quality metrics of the grayscale filter application supplementary simulation based on the pictures of \cite{datasetfig}.}
	\scalebox{0.9}{
	\begin{tabular}{|c|c|c|c|}
		\hline
		8-bit Serial Approximate RCA & PSNR (dB) & SSIM & MSSIM \\ \hline
		\multicolumn{4}{|c|}{$1^{st}~scenario:~the~four~MSBs~are~exact,~and~the~four~LSBs~are~approximate.$} \\ \hline
		FAFA1 and FAFA2 & 41.998 & 0.976 & 0.99 \\ \hline
		\multicolumn{4}{|c|}{$2^{nd}~scenario:~the~three~MSBs~are~exact,~and~the~five~LSBs~are~approximate.$} \\ \hline
		FAFA1 and FAFA2 & 36.034 & 0.93 & 0.97 \\ \hline
	\end{tabular}}
	\label{taba3}
\end{table}

\begin{table}[!]
	\centering
	\caption{Average Image quality metrics of the average pooling application supplementary simulation based on the pictures of \cite{datasetfig}.}
	\scalebox{0.9}{
	\begin{tabular}{|c|c|c|c|}
		\hline
		8-bit Serial Approximate RCA & PSNR (dB) & SSIM & MSSIM \\ \hline
		\multicolumn{4}{|c|}{$1^{st}~scenario:~the~four~MSBs~are~exact,~and~the~four~LSBs~are~approximate.$} \\ \hline
		FAFA1 and FAFA2 & 40.407 & 0.973 & 0.974 \\ \hline
		\multicolumn{4}{|c|}{$2^{nd}~scenario:~the~three~MSBs~are~exact,~and~the~five~LSBs~are~approximate.$} \\ \hline
		FAFA1 and FAFA2 & 33.858 & 0.912 & 0.915 \\ \hline
	\end{tabular}}
	\label{taba4}
\end{table}

\section*{Statements \& Declarations}
\subsection*{Funding}
No funds, grants, or other support was received.

\subsection*{Competing interests}
The authors declare no competing interests.

\subsection*{Author Contributions}
\noindent \textbf{Seyed Erfan Fatemieh}: Conceptualization, Methodology, Software, Validation, Formal Analysis, Investigation, Writing - Original Draft, Writing – Review \& Editing, Visualization.

\noindent \textbf{Samane Asgari}: Conceptualization, Methodology, Software, Validation, Formal Analysis, Investigation, Writing - Original Draft, Writing - Review \& Editing, Visualization. 

\noindent \textbf{Mohammad Reza Reshadinezhad}: Validation, Formal Analysis, Investigation, Resources, Writing - Review \& Editing, Supervision, Project Administration.  

\subsection*{Data Availability Statement}
All data have been generated through the methods described in this paper.

\bibliographystyle{ieeetr} 
\bibliography{Manuscript}

\end{document}